\documentclass[reprint,
 amsmath,amssymb,
 aps,
]{revtex4-2}

\usepackage{graphicx}
\usepackage{dcolumn}
\usepackage{bm}
\usepackage{xcolor}
\usepackage{mathptmx}
\usepackage{graphicx}
\usepackage{dcolumn}
 
\usepackage{bm}
\usepackage{amssymb}
\usepackage{comment}
\usepackage{hyperref}

\begin{document}

\title{Synthesis and characterization of space-time light sheets: A tutorial}
\author{Miguel A. Romer$^{1}$}
\author{Layton A. Hall$^{1}$}
\author{Ayman F. Abouraddy$^{1,*}$}
\affiliation{$^{1}$CREOL, The College of Optics \& Photonics, University of Central~Florida, Orlando, FL 32816, USA}
\affiliation{$^*$Corresponding author: raddy@creol.ucf.edu}

\begin{abstract}
Space-time wave packets (STWPs) are a new class of pulsed optical beams with many unique and intriguing attributes, including propagation invariance and tunable group velocity in linear optical media. STWPs are a form of spatiotemporally structured light, so their synthesis poses challenges that are not shared by conventional monochromatic structured light fields. We present here a detailed description of the synthesis of STWPs that are localized along one transverse dimension and uniform along the other; i.e., space-time light sheets. We also describe the main characterization schemes needed for benchmarking the unique properties of space-time light sheets.
\end{abstract}


\maketitle

\section{Introduction}

The structuring of light in the spatial and temporal domains has taken major strides over the past few decades \cite{Yessenov22AOP}. In the \textit{spatial} domain, a host of new modal sets have been discovered and explored, including orbital angular momentum modes \cite{Allen92PRA,Allen03Book,Yao11AOP}, Bessel beams \cite{Durnin87PRL,McGloin05CP}, Airy beams \cite{Siviloglou07OL,Efremidis19Optica}, to name just a few. These are typically monochromatic beams that are sculpted spatially to conform to a particular structure with desired attributes \cite{Forbes16AOP,Forbes21NP}. Moreover, earlier efforts in the field of Fourier Optics provided a boon for spatial structuring of optical fields \cite{Flannery89ProcIEEE,GoodmanBook05,SalehBook07}. In the \textit{temporal} domain, ultrafast pulse shaping in the picosecond and femtosecond regimes via spectral phase and/or amplitude modulation \cite{Weiner88JOSAB,Weiner00RSI} has enabled new applications in nonlinear optics and optical communications \cite{Weiner09Book}. However, the research communities that have focused on \textit{spatially} structured beams and \textit{temporally} structured pulses have remained mostly independent of each other, maintaining distinct modulation techniques and application spaces.

Recently, new vistas have been opened by exploring \textit{spatiotemporally} structured light. By this we mean optical fields in which the spatial and temporal degrees-of-freedom (DoFs) are modulated \textit{jointly}, so that the field is no longer separable with respect to space and time as is the case for conventional pulsed beams \cite{Yessenov22AOP}. It is well-known that the spatial and temporal DoFs naturally become coupled after traversing a variety of generic optical components \cite{Akturk05OE}, or even upon free propagation when the pulsewidth is very short and the beam size small (e.g., focusing of ultrashort pulses \cite{Zhu05OE,SalehBook07}). Such coupling is usually viewed as a nuisance to be ameliorated \cite{Wikmark19PNAS,Dorrer19IEEEJSTQE,Jeandet22OE}. Nevertheless, spatiotemporal coupling when intentionally introduced into an optical field can produce novel and desirable propagation behaviors. This concept was exploited in the development of propagation-invariant pulsed beams dating back to Brittingham's focus-wave modes (FWMs) \cite{Brittingham83JAP}, and subsequently the X-wave \cite{Lu92IEEEa} realized in optics by P.~Saari \cite{Saari97PRL}. However, these classes of pulsed beams (or wave packets) unfortunately require ultra-broad bandwidths to be realized, and their propagation characteristics (specifically, their group velocity) deviate from those of conventional pulsed beams only in the non-paraxial regime \cite{Yessenov22AOP}. Consequently, despite the intriguing characteristics of FWMs and X-waves, no optical applications have benefited from them to date \cite{Turunen10PO}.

This state of affairs has changed dramatically over the past few years with the emergence of so-called `space-time wave packets' (STWPs) \cite{Kondakci16OE,Parker16OE,Wong17ACSP2,Efremidis17OL,Porras17OL,Kondakci17NP,Kondakci18PRL,Guo21Light,Bejot21ACSP,Bejot22ACSP,Stefanska23ACSP,Ramsey23PRA} (see the review in \cite{Yessenov22AOP}), which are propagation-invariant pulsed beams in linear media \cite{Turunen10PO,FigueroaBook14}. Uniquely, STWPs can be realized with small bandwidths in the paraxial regime -- even when their characteristics deviate drastically from those of a conventional pulsed beam \cite{Yessenov22AOP}. To date, a host of unique and useful features of STWPs have been predicted and verified experimentally, including:
\begin{enumerate}
\item Propagation invariance: STWPs are propagation invariant (diffraction-free and dispersion-free) in free space \cite{Bhaduri18OE,Bhaduri18OL}, in linear non-dispersive media \cite{Bhaduri19Optica,Bhaduri20NP}, and even in dispersive media \cite{Sonajalg96OL,Sonajalg97OL,Porras03PRE2,Malaguti08OL,Malaguti09PRA,Yessenov22OLdispersion,He22LPR,Hall23LPR,Hall23Normal,Hall23NPhys,Hall24XtoO}. The propagation distance has been recently extended to the kilometer range \cite{Hall22arxiv1km}, which suggests potential applications in free-space optical communications, remote sensing, and directed energy.  
\item Tunable group velocity: The group velocity of an STWP is determined by its spatiotemporal structure rather than being dictated by the properties of the medium. As such, the STWP group velocity $\widetilde{v}$ is readily tunable \cite{Salo01JOA,Wong17ACSP2,Efremidis17OL} in free space across the subluminal ($\widetilde{v}\!<\!c$, where $c$ is the speed of light in vacuum), superluminal ($\widetilde{v}\!>\!c$), or even negative-$\widetilde{v}$ ($\widetilde{v}\!<\!0$) regimes \cite{Kondakci19NC,Yessenov19OE}. Similar control can be exercised over $\widetilde{v}$ in linear media \cite{Bhaduri19Optica,Bhaduri20NP}.
\item Anomalous refraction: Because refraction of optical fields at a planar interface involves conservation of both energy and the momentum component parallel to the interface, the spatiotemporal structure of the STWP is modified upon transmission. Consequently, the STWP group velocity changes with refraction \cite{Bhaduri20NP}, leading to a variety of never-before-seen phenomena at normal incidence: group-velocity \textit{invariance} (an STWP that traverses the interface between media with different refractive indices without change in $\widetilde{v}$); \textit{anomalous} refraction (increase in $\widetilde{v}$ upon transmission to a higher-index medium; group-velocity \textit{inversion} (a switch in the sign of $\widetilde{v}$ for an STWP after traversing the interface) \cite{Yessenov21JOSAAI,Motz21JOSAAII}. At \textit{oblique} incidence, other refractive phenomena emerge, such as isochronous STWPs (STWPs that incur the same group delay after traversing a dielectric slab regardless of incident angle) \cite{Motz21OL} and blind clock synchronization (simultaneous arrival of an STWP broadcast to multiple receivers at the same depth but different locations beyond the interface)  \cite{Yessenov21JOSAAI,Yessenov21JOSAAIII}.
\item Additionally, a single feature of the STWP field can be singled out and made to vary controllably along the propagation axis. For example, the spectrum at the beam center can be made to change in an almost arbitrary fashion along the propagation axis, a feature we have called `axial spectral encoding' \cite{Motz21PRA}. One may also introduce an axially varying group velocity (i.e., an axially accelerating wave packet) \cite{Clerici08OE,ValtnaLukner09OE,Yessenov20PRL2,Li20SR,Li20CP,Li21CP,Hall22OLArbAccel}.
\end{enumerate}

No currently known laser directly emits STWPs; rather, they are synthesized starting from generic laser pulses \cite{Yessenov19OPN,Yessenov22AOP}. Nevertheless, the useful properties of STWPs and the ease by which they can be synthesized make them primed for deployment in a variety of applications. The goal of this tutorial is to present a detailed description of how to synthesize an STWP starting from a generic pulsed laser beam. An additional goal is to develop a systematic characterization protocol to benchmark the properties of STWPs. The early successful demonstrations of STWPs made use of fields localized spatially along only one transverse dimension but uniform along the other; i.e., fields in the form of a light sheet. Only recently have STWPs been synthesized that are localized in both transverse dimensions \cite{Pang21OL,Pang22OE,Yessenov22NC,Yessenov22OL3D,Zou22OL,Minoofar22OE}, and developments in this area are on-going. We therefore focus in this tutorial on the synthesis and characterization of STWPs in the form of light sheets with the aim of enabling a newcomer to readily establish their own experimental testbed.

\section{General theoretical formulation}

\subsection{Conventional pulsed beams}

Conventional pulsed laser beams are typically separable with respect to their spatial and temporal DoFs. When we restrict our attention to one transverse dimension $x$ in the paraxial regime, the field can be expressed as $E(x,z;t)\!=\!e^{i(k_{\mathrm{o}}z-\omega_{\mathrm{o}}t)}\psi(x,z;t)$, where $\omega_{\mathrm{o}}$ is a temporal carrier frequency, $k_{\mathrm{o}}\!=\!\omega_{\mathrm{o}}/c$ is the associated wave number, and the angular spectrum of the slowly varying envelope $\psi(x,z;t)$ is written as follows:
\begin{equation}\label{Eq:ConventionalEnvelope}
\psi(x,z;t)\!=\!\iint\!dk_{x}d\Omega\;\widetilde{\psi}(k_{x},\Omega)e^{i\{k_{x}x+(k_{z}-k_{\mathrm{o}})z-\Omega t\}},
\end{equation}
where $\Omega\!=\!\omega-\omega_{\mathrm{o}}$, $\omega$ is the temporal frequency, $k_{z}\!=\!\sqrt{(\tfrac{\omega}{c})^{2}-k_{x}^{2}}$, and the spatiotemporal spectrum $\widetilde{\psi}(k_{x},\Omega)$ is the Fourier transform of $\psi(x,0;t)$. The intensity of the wave packet is:
\begin{equation}
I(x,z;t)=|\psi(x,z;t)|^{2}.
\end{equation}
It is crucial to appreciate that $k_{x}$ and $\Omega$ are, in principle, independent variables. The integral in Eq.~\ref{Eq:ConventionalEnvelope} could have equivalently been expressed in terms of the variables $k_{z}$ and $\Omega$ rather than $k_{x}$ and $\Omega$. We choose to use the latter because $k_{x}$ and $\Omega$ can be directly manipulated experimentally.

In addition to the spatiotemporal spectrum in Eq.~\ref{Eq:ConventionalEnvelope}, we define other `marginal' distributions for the spatial and temporal DoFs separately.
\begin{enumerate}
    \item The pulse profile at the beam center $I_{t}(t)\!=\!|\psi_{t}(t)|^{2}\!=\!I(0,0;t)$, where:
    \begin{equation}
    \psi_{t}(t)=\psi(x=0,z=0;t)=\int\!d\Omega\widetilde{\psi}_{t}(\Omega)e^{-i\Omega t},
    \end{equation}
    and $\widetilde{\psi}_{t}(\Omega)\!=\!\int\!dk_{x}\widetilde{\psi}(k_{x},\Omega)$. Note that $\widetilde{\psi}_{t}(\Omega)$ results from the integration of the complex spectral amplitude $\widetilde{\psi}(k_{x},\Omega)$ over $k_{x}$ and \textit{not} from integration over the spectral intensity $|\widetilde{\psi}(k_{x},\Omega)|^{2}$. Therefore, one may have $\psi_{t}(t)\!=\!0$ for all $t$ if $\widetilde{\psi}(k_{x},\Omega)$ has a phase structure along $k_{x}$ \cite{Wong21,Hall23NPhys,Diouf23NPhys}.   
    \item The beam profile at the pulse center $I_{x}(x)\!=\!|\psi_{x}(x)|^{2}\!=\!I(x,0;0)$, where:
     \begin{equation}
    \psi_{x}(x)=\psi(x,z=0;t=0)=\int\!dk_{x}\widetilde{\psi}_{x}(k_{x})e^{ik_{x}x},
    \end{equation}
    and $\widetilde{\psi}_{x}(x)\!=\!\int\!d\Omega\,\widetilde{\psi}(k_{x},\Omega)$. Again, $\widetilde{\psi}_{x}(k_{x})$ results from integrating the complex spectral amplitude $\widetilde{\psi}(k_{x},\Omega)$ over $\Omega$. Therefore, one may have $\psi_{x}(x)\!=\!0$ for all $x$ if $\widetilde{\psi}(k_{x},\Omega)$ has a phase structure along $\Omega$ \cite{Kondakci18PRL,Yessenov19Optica,Wong21}. 
    \item The time-averaged spatial intensity profile, which is given by:
    \begin{equation}
    I(x)=\int\!dt\;I(x,0;t).
    \end{equation}
    This is the intensity distribution recorded by a `slow' detector (such as a CCD camera) placed at a given axial plane.
\end{enumerate}

In conventional pulsed beams, the spatiotemporal spectrum is usually separable with respect to the spatial and temporal DoFs, $\widetilde{\psi}(k_{x},\Omega)\!\approx\!\widetilde{\psi}_{x}(k_{x})\widetilde{\psi}_{t}(\Omega)$, in which case the temporal component $\widetilde{\psi}_{t}(\Omega)$ of the spectrum determines the temporal profile $\psi_{t}(t)$ without any impact from the spatial DoF. Moreover, the spatial beam profile $\psi_{x}(x)$ is determined by the spatial component $\widetilde{\psi}_{x}(k_{x})$  of the spatiotemporal spectrum without any impact from the temporal DoF. In other words, the initial spatiotemporal profile is separable with respect to space and time, $\psi(x,0;t)\!\approx\!\psi_{x}(x)\psi_{t}(t)$ and $I(x,0;t)\!\approx\!I_{x}(x)I_{t}(t)$. Consequently, in the case of a separable field we have $I(x)\!=\!I_{x}(x)$, so that the transverse spatial width of the beam is the same whether recorded by a slow detector or acquired at the center of the pulse from time-resolved measurements. The situation is different when the spectrum is \textit{non-separable}, as is the case for STWPs. Indeed, in the case of STWPs, the width of the time-averaged intensity $I(x)$ and that of the beam profile at the pulse center $I_{x}(x)$ differ by a factor of 2 \cite{Yessenov19Optica}, as we demonstrate below.

\subsection{Spectral representation on the surface of the light-cone}

\begin{figure}[t]
    \centering
    \includegraphics[width=86mm]{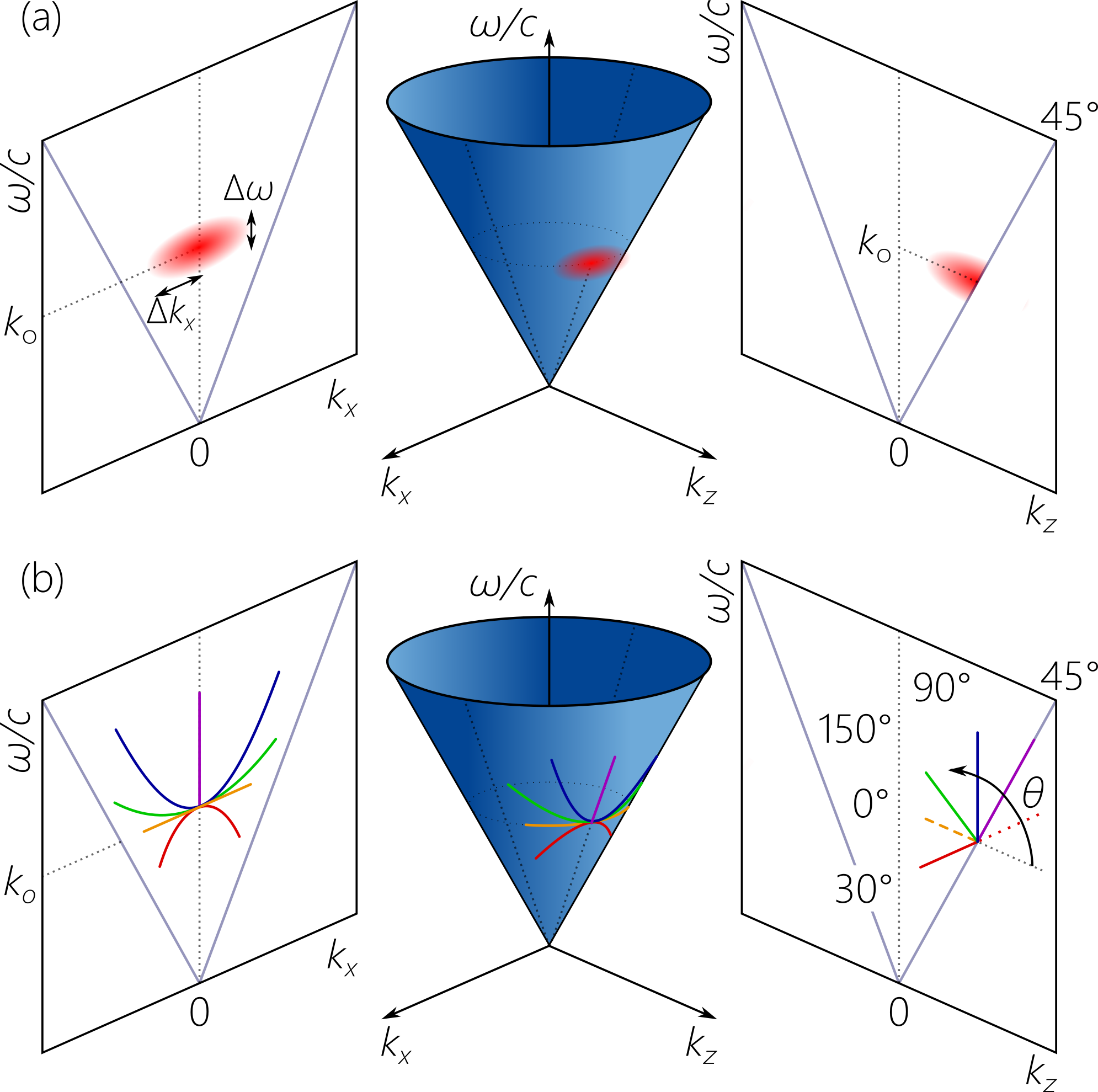}
    \caption{(a) Spectral support for a conventional pulsed beam on the surface of the light-cone in $(k_{x},k_{z},\tfrac{\omega}{c})$-space, along with the spectral projections onto the $(k_{x},\tfrac{\omega}{c})$ and $(k_{z},\tfrac{\omega}{c})$ planes. The spatial and temporal bandwidths are $\Delta k_{x}$ and $\Delta\omega$, respectively. (b) Spectral support for STWPs with spectral tilt angles $\theta\!=\!30^{\circ},45^{\circ},90^{\circ}$, and $150^{\circ}$ on the light-cone surface. The spectral projections onto the $(k_{x},\tfrac{\omega}{c})$- plane are conic sections that are approximately parabolas in the vicinity of $k_{x}\!=\!0$. The spectral projections onto the $(k_{z},\tfrac{\omega}{c})$-plane are straight lines making an angle $\theta$ with the $k_{z}$-axis and passing through the point $(k_{z},\tfrac{\omega}{c})\!=\!(k_{\mathrm{o}},k_{\mathrm{o}})$. All the dotted lines correspond to light-lines $|k_{x}|\!=\!\tfrac{\omega}{c}$ or $k_{z}\!=\!\tfrac{\omega}{c}$.}
    \label{Fig:LightCones}
\end{figure}

A useful theoretical tool for understanding and visualizing the characteristics of STWPs is the representation of the spatiotemporal spectrum on the surface of the free-space light-cone $k_{x}^{2}+k_{z}^{2}\!=\!(\tfrac{\omega}{c})^{2}$ [Fig.~\ref{Fig:LightCones}]. Any monochromatic plane-wave $e^{i(k_{x}x+k_{z}z-\omega t)}$ is represented by a point on the surface of the light-cone in $(k_{x},k_{z},\tfrac{\omega}{c})$-space. Therefore, for any physically realizable optical field that can be expanded by an angular spectrum as in Eq.~\ref{Eq:ConventionalEnvelope}, the spatiotemporal spectrum $\widetilde{\psi}(k_{x},\Omega)$ corresponds to some domain on the light-cone surface \cite{Donnelly93PRSLA}.

For example, in the special case of a \textit{monochromatic} beam where $\widetilde{\psi}(k_{x},\Omega)\!\rightarrow\!\widetilde{\psi}(k_{x})\delta(\Omega)$, the spatial spectrum corresponds to a 1D trajectory on the light-cone; namely, a portion of the circle at the intersection of the light-cone with an iso-frequency plane $\omega\!=\!\omega_{\mathrm{o}}$ \cite{Yessenov22AOP}. In the more general case of a conventional pulsed beam, the spatiotemporal spectrum $\widetilde{\psi}(k_{x},\Omega)$ corresponds to a 2D domain on the light-cone surface. This reflects the fact that the pulsed beam has finite spatial \textit{and} temporal bandwidths $\Delta k_{x}$ and $\Delta\omega$, respectively, because of the finite spatial extent of the transverse beam profile and the finite temporal extent of the pulse profile [Fig.~\ref{Fig:LightCones}(a)].

\subsection{Formulation of STWPs}

Whereas the spatial and temporal spectral DoFs are typically separable for a conventional pulsed beam, the unique characteristics of STWPs stem, in contrast, from their non-separability. Specifically, the spatiotemporal spectrum for an ideal STWP is a 1D function $\widetilde{\psi}(k_{x},\Omega)\!\rightarrow\!\widetilde{\psi}(\Omega)\delta(k_{x}-k_{x}(\Omega))$, in which each spatial frequency $k_{x}$ is associated with a single temporal frequency $\Omega$. In other words, $k_{x}$ and $\Omega$ are no longer independent variables. Consequently, the field $E(x,z;t)\!=\!e^{i(k_{\mathrm{o}}z-\omega_{\mathrm{o}}t)}\psi(x,z;t)$ has an envelope whose angular spectrum is expressed as follows:
\begin{equation}\label{Eq:STWPEnvelope1}
\psi(x,z;t)=\int\!d\Omega\widetilde{\psi}(\Omega)e^{i\{k_{x}x+(k_{z}-k_{\mathrm{o}})z-\Omega t\}},
\end{equation}
where neither $k_{z}$ nor $k_{x}$ are independent variables, and they instead both depend on $\Omega$.

In general, propagation-invariance of an STWP requires that the tight association between the spatial and temporal frequencies satisfies the following constraint \cite{FigueroaBook14,Yessenov22AOP}:
\begin{equation}\label{Eq:SpectralPlane}
\omega-\omega_{\mathrm{o}}=(k_{z}-k_{\mathrm{o}})\widetilde{v},
\end{equation}
in which case the envelope in Eq.~\ref{Eq:STWPEnvelope1} takes the form
\begin{equation}\label{Eq:STWPEnvelope}
\psi(x,z;t)=\int\!d\Omega\widetilde{\psi}(\Omega)e^{i\{k_{x}x-\Omega(t-z/\widetilde{v})\}}=\psi(x,0;t-z/\widetilde{v}).
\end{equation}
This envelope represents a wave packet traveling rigidly in free space -- without diffraction or dispersion -- at a group velocity $\widetilde{v}$. In principle, $\widetilde{v}$ can take on any value \cite{Salo01JOA}, so that the second most salient characteristic of STWPs after their propagation invariance is that their group velocity is readily tunable in free space \cite{Kondakci19NC,Bhaduri20NP} (or in any linear medium).

The marginal spectra can be obtained using the formalism presented above: $\widetilde{\psi}_{t}(\Omega)\!=\!\int\!dk_{x}\;\widetilde{\psi}(\Omega)\delta(k_{x}-k_{x}(\Omega))\!=\!\widetilde{\psi}(\Omega)$ and $\widetilde{\psi}_{x}(k_{x})\!=\!\int\!d\Omega\;\widetilde{\psi}(\Omega)\delta(k_{x}-k_{x}(\Omega))$. The latter expression can be evaluated after positing a specific spectral model. It is clear that $\widetilde{\psi}(k_{x},\Omega)\!\neq\!\widetilde{\psi}_{x}(k_{x})\widetilde{\psi}_{t}(\Omega)$, so that the spatiotemporal spectrum cannot be reconstructed from the marginal spectra. Moreover, the spatiotemporal profile is \textit{not} separable with respect to $x$ and $t$: $\psi(x,0;t)\!\neq\!\psi_{x}(x)\psi_{t}(t)$. The marginal profiles $\psi_{x}(x)$ and $\psi_{t}(t)$ are obtained by taking the spatial and temporal Fourier transforms of $\widetilde{\psi}_{x}(k_{x})$ and $\widetilde{\psi}_{t}(\Omega)$, respectively.

Because of the non-separability of STWPs with respect to their spatial and temporal DoFs, a new structure emerges in the time-averaged intensity profile. Rewriting the spatiotemporal spectrum as $\widetilde{\psi}(k_{x},\Omega)\!=\!\widetilde{\psi}(k_{x})\delta(\Omega-\Omega(k_{x}))$ for convenience, it can be readily shown that the axial evolution of the time-averaged intensity profile $I(x,z)\!=\!\int\!dt\,|\psi(x,z;t)|^{2}$ is given by:
\begin{equation}\label{eq:TimeAveraged}
I(x,z)=\int\!dk_{x}\;|\widetilde{\psi}(k_{x})|^{2}+\mathrm{Re}\int\!dk_{x}\widetilde{\psi}(k_{x})\widetilde{\psi}^{*}(-k_{x})e^{i2k_{x}x}.
\end{equation}
Several conclusions can be readily drawn from this result:
\begin{enumerate}
    \item The time-averaged intensity for an \textit{ideal} STWP is independent of $z$, $I(x,z)\!=\!I(x,0)$, and is thus diffraction-free.
    \item The transverse structure of $I(x,0)$ takes the form of a constant pedestal (the first term in Eq.~\ref{eq:TimeAveraged}) atop of which is a spatially localized feature (the second term in Eq.~\ref{eq:TimeAveraged}).
    \item Because of the additional factor-of-2 in the exponent $e^{i2k_{x}x}$, the width of this spatial feature is half that expected from the width of the spatial spectrum. We thus expect a discrepancy between this spatial width and that of the beam at the pulse center $I_{x}(x)\!=\!|\psi_{x}(x)|^{2}$. This characteristic is absent from conventional pulsed optical beams.
\end{enumerate}

\subsection{STWP spectra on the light-cone surface}

If we write the STWP group velocity as $\widetilde{v}\!=\!c\tan{\theta}$, then the representation of its spatiotemporal spectrum on the light-cone surface takes on a clear geometric meaning. The spectral support is a 1D trajectory at the intersection of the light-cone surface with the tilted spectral plane given by Eq.~\ref{Eq:SpectralPlane}. This plane is parallel to the $k_{x}$-axis, makes an angle $\theta$ with the $k_{z}$-axis, and passes through the point $(k_{x},k_{z},\tfrac{\omega}{c})\!=\!(0,k_{\mathrm{o}},k_{\mathrm{o}})$. We thus refer to $\theta$ as the spectral tilt angle. The intersection of such a plane with a cone is a conic section whose type depends on $\theta$ \cite{Yessenov19PRA}: an ellipse when $0\!<\!\theta\!<\!45^{\circ}$ or $135\!<\!\theta\!<\!180^{\circ}$, a line when $\theta\!=\!45^{\circ}$, a hyperbola when $45\!<\!\theta\!<\!135^{\circ}$, or a parabola when $\theta\!=\!135^{\circ}$. We refer to the regime where $\widetilde{v}\!<\!c$ as subluminal ($0\!<\!\theta\!<\!45^{\circ}$), where $\widetilde{v}\!>\!c$ as superluminal ($45\!<\!\theta\!<\!90^{\circ}$), and where $\widetilde{v}\!<\!0$ as negative-$\widetilde{v}$ ($90\!<\!\theta\!<\!180^{\circ}$) [Fig.~\ref{Fig:LightCones}(b)].

In all cases, the projection of this spectrum onto the $(k_{x},\tfrac{\omega}{c})$-plane is a conic section, whereas its projection onto the $(k_{z},\tfrac{\omega}{c})$-plane is a straight-line passing through the point $(k_{z},\tfrac{\omega}{c})\!=\!(k_{\mathrm{o}},k_{\mathrm{o}})$ and making an angle $\theta$ with the $k_{z}$-axis. Although the nature of the conic section representing the STWP spectrum varies with $\theta$, we are typically interested in the paraxial ($\Delta k_{x}\!\ll\!k_{\mathrm{o}}$) narrowband ($\Delta\omega\!\ll\!\omega_{\mathrm{o}}$) regime, whereupon the conic section in the vicinity of $k_{x}\!=\!0$ can be approximated by the parabola \cite{Kondakci17NP,Yessenov22AOP}:
\begin{equation}\label{Eq:Parabola}
\frac{\Omega}{\omega_{\mathrm{o}}}\approx\frac{1}{1-\cot{\theta}}\;\;\frac{k_{x}^{2}}{2k_{\mathrm{o}}^{2}}.
\end{equation}
When $\theta\!<\!45^{\circ}$ ($\cot{\theta}\!>\!1$), then $\Omega\!<\!0$, which reflects the fact that $\omega\!<\!\omega_{\mathrm{o}}$ in the subluminal regime [Fig.~\ref{Fig:LightCones}(b)]. Alternatively, when $45^{\circ}\!<\!\theta\!<\!180^{\circ}$ ($\cot{\theta}\!<\!1$), then $\Omega\!>\!0$, which reflects the fact that $\omega\!>\!\omega_{\mathrm{o}}$ in the superluminal and negative-$\widetilde{v}$ regimes [Fig.~\ref{Fig:LightCones}(b)].

Because each spatial frequency $k_{x}$ is associated with a single temporal frequency $\omega$ (Eq.~\ref{Eq:Parabola}), STWPs are endowed with `angular dispersion' \cite{Torres10AOP,Fulop10Review}; that is, each frequency $\omega$ travels at a different angle $\varphi(\omega)$ with the $z$-axis, where $k_{x}(\omega)\!=\!\tfrac{\omega}{c}\sin\{\varphi(\omega)\}$. The phenomenon of angular dispersion has been extensively studied since the pioneering work of Newton \cite{Sabra81Book}. In conventional angular dispersion, it is always assumed that $\varphi(\omega)$ is a differentiable function with respect to $\omega$ at $\omega\!=\!\omega_{\mathrm{o}}$, so that $\varphi(\omega)$ can be expanded in a Taylor series around $\omega_{\mathrm{o}}$. In contrast, it is clear that the angular dispersion undergirding an STWP (Eq.~\ref{Eq:Parabola}) is \textit{non-differentiable} because $\varphi(\omega)\!\propto\!\sqrt{\omega-\omega_{\mathrm{o}}}$, whereupon the derivative of such a function is \textit{not} defined at $\omega\!=\!\omega_{\mathrm{o}}$. We have recently demonstrated that the non-differentiable angular dispersion intrinsic to STWPs underlines the departure of the characteristics of STWPs from those of all other fields endowed with conventional differentiable angular dispersion \cite{Hall21OLViolate,Yessenov21ACSP,Hall21OLNormalGVD,Hall22OEConsequences,Hall22JOSAA}.  

\section{Synthesis and characterization of space-time light sheets}

\begin{figure}[t]
    \centering
    \includegraphics[width=86mm]{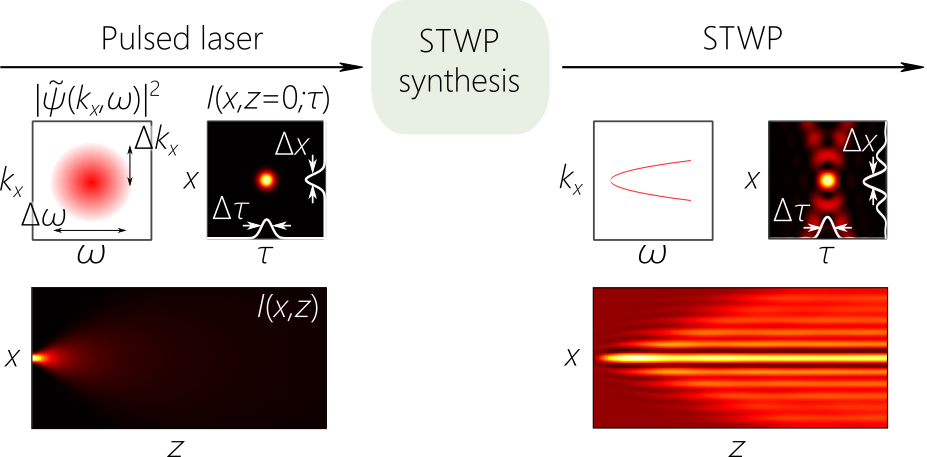}
    \caption{Synthesizing an STWP from a generic pulsed laser beam. The input to the STWP synthesis system is a conventional pulsed laser beam with separable spatiotemporal structure. The synthesis system produces a propagation-invariant STWP with non-separable spatiotemporal structure. For the input and output fields, we plot the spatiotemporal spectrum $|\widetilde{\psi}(k_{x},\omega)|^{2}$, spatiotemporal intensity profile $I(x,0;\tau)$, and the axial evolution of the time-averaged intensity $I(x,z)$. The white curves in the plots of $I(x,0;\tau)$ are the marginal envelopes $I_{x}(x)$ and $I_{t}(t)$.}
    \label{Fig:BasicConception}
\end{figure}

We first outline conceptually the general procedures we have employed to synthesize STWPs in the form of light sheets and to benchmark their characteristics, before moving on to a more detailed description of the experimental apparatus.

\subsection{Synthesis of ST light sheets}

To synthesize an STWP, we start with a generic pulsed laser beam and produce the STWP through spatiotemporal spectral phase modulation, as illustrated in Fig.~\ref{Fig:BasicConception}. The starting point is a generic pulsed laser beam with separable spatial and temporal DoFs $\widetilde{\psi}(k_{x},\Omega)\!\approx\!\widetilde{\psi}_{t}(\Omega)\widetilde{\psi}_{x}(k_{x})$. Such a field undergoes diffractive spatial spreading with propagation. From this generic picosecond or femtosecond pulse, an STWP synthesizer modifies the structure of the spatiotemporal spectrum and assigns each temporal frequency $\Omega$ to a prescribed pair of spatial frequencies $\pm k_{x}$ according to Eq.~\ref{Eq:Parabola}. The spatiotemporal structure of the field is no longer separable, and the field is instead usually takes an X-shaped structure in space and time \cite{FigueroaBook14,Yessenov22AOP} (an exception is the O-wave structure of STWPs in the presence of anomalous GVD \cite{Malaguti08OL,Hall23NPhys,Diouf23NPhys}). Such an STWP synthesizer can implement -- in principle -- any spectral tilt angle $\theta$, and hence realize a propagation-invariant STWP of group velocity $\widetilde{v}\!=\!c\tan{\theta}$ \cite{Kondakci17NP}.

\begin{figure}[t]
    \centering
    \includegraphics[width=86mm]{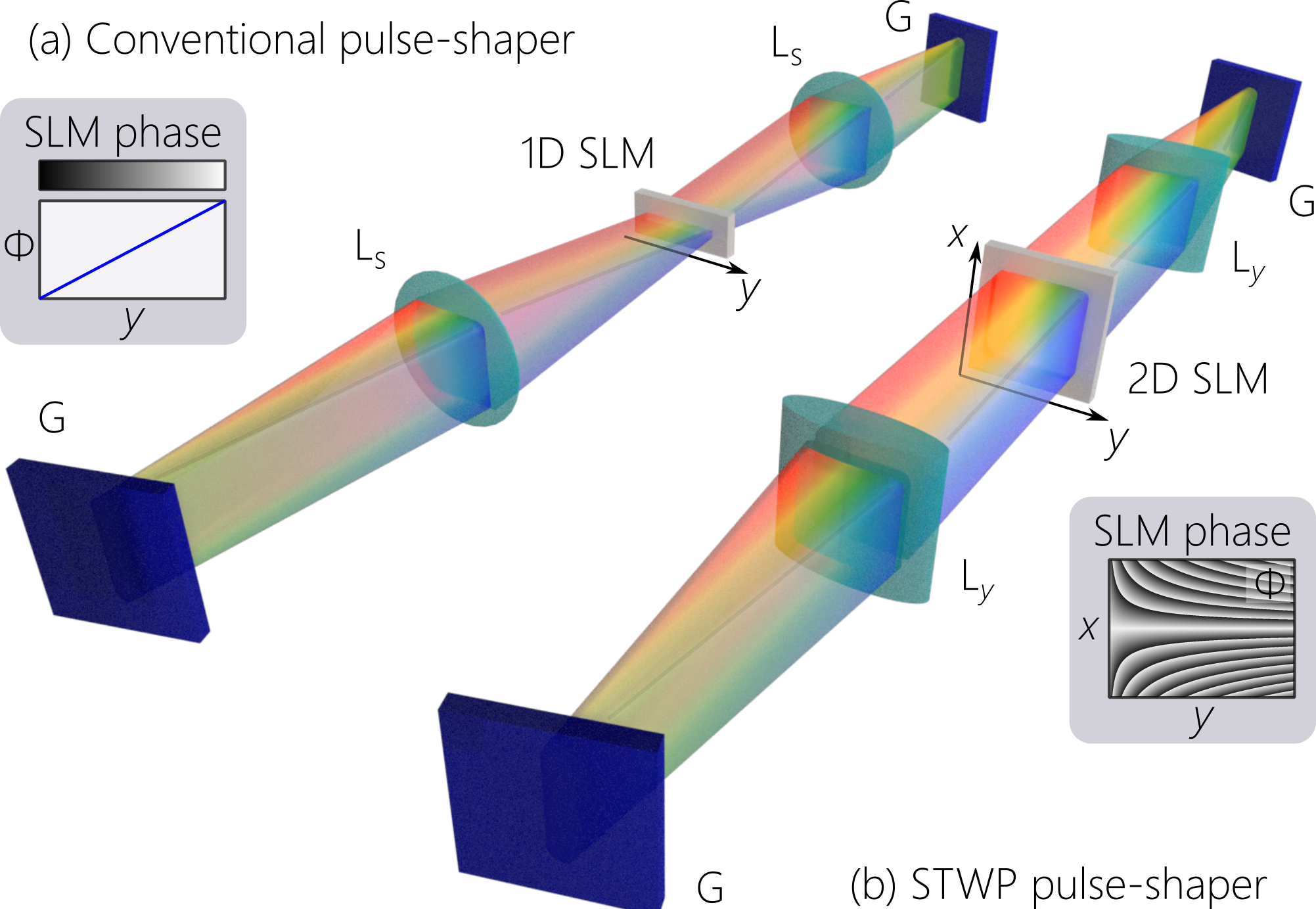}
    \caption{Comparison between a typical setup for conventional ultrafast pulse-shaping via spectral-phase modulation and an STWP synthesizer. (a) Schematic of a conventional $4f$ pulse-shaper \cite{Weiner00RSI,Weiner09Book} containing a 1D SLM. The inset depicts an example of the SLM phase profile $\Phi(y)$ to produce a group delay. G: Diffraction grating; L$_{\mathrm{s}}$: spherical lens. (b) Schematic of an STWP synthesizer containing a 2D SLM. L$_{y}$: Cylindrical lens focusing along the $y$-axis (the direction along which the spectrum is spread). Inset shows the 2D SLM phase profile $\Phi(x,y)$ to produce a subluminal STWP (see Fig.~\ref{Fig:SLMPhasePatterns}).}
    \label{Fig:ComparisonOfPulseShaperAndSTWP}
\end{figure}

\subsubsection{Temporal pulse shaping}

As a first step to understanding how to construct an STWP synthesizer, consider the setup for the well-established spectral-phase modulation technique for ultrafast pulse shaping as depicted in Fig.~\ref{Fig:ComparisonOfPulseShaperAndSTWP}(a). It makes use of a grating to spatially resolve the spectrum, which is focused by a spherical lens to a 1D linear spatial light modulator (SLM; typically phase-only, although phase and amplitude modulation can be used) depicted in transmission mode. Each wavelength (or narrow bandwidth $\delta\lambda$) is mapped to a pixel of the SLM, and the modulated spectrum is returned to a symmetrically placed grating via a second spherical lens. The modulated pulse is reconstituted at the second grating after the wavelengths are overlapped. The spatial profile of the field after this pulse-shaper is the same as at its input; only the temporal pulse profile changes.

\subsubsection{Spatiotemporal spectral shaping}

Previous approaches followed to synthesize propagation-invariant wave packets include making use of Bessel-beam generators and related techniques \cite{Saari97PRL,Alexeev02PRL,Valtna07OC,Bonaretti09OE,Bowlan09OL} nonlinear optical processes \cite{DiTrapani03PRL,Faccio06PRL,Faccio07OE}, or spatiotemporal spectral amplitude filtering \cite{Dallaire09OE,Jedrkiewicz13OE}. To synthesize an STWP, we have relied in our work on spatiotemporal spectral phase modulation achieved using the same overall strategy of the pulse shaper in Fig.~\ref{Fig:ComparisonOfPulseShaperAndSTWP}(a) after a few crucial modifications. First, the 1D SLM is replaced with a 2D SLM [Fig.~\ref{Fig:ComparisonOfPulseShaperAndSTWP}(b)]. The spectrum of the incident pulse is spatially resolved by the first grating and focused to the SLM via a \textit{cylindrical} lens (rather than a spherical lens). Consequently, each wavelength occupies a column of the SLM rather than only a pixel. One may now impart a phase distribution along this column to \textit{spatially} modulate each wavelength independently of all other wavelengths incident on the SLM. In other words, each wavelength is now associated with a distinct spatial phase distribution that can differ from that associated with all other wavelengths. The spatially modulated, spectrally resolved wave front is directed symmetrically via a cylindrical lens to a second grating, whereupon the STWP is constituted. Both the spatial and the temporal profiles of the output field differ from those at the input. Crucially, the spatial and temporal DoFs are no longer separable. Careful design of the 2D spatial phase distribution imparted by the SLM to the field is critical for producing an STWP with the targeted spectral tilt angle $\theta$.

\subsection{Characterization of STWPs}

Given the fundamental differences between conventional pulsed beams and STWPs \cite{Hall22OEConsequences}, which have their origin in the spatiotemporal separability of the former and non-separability of the latter, it is critical to clearly benchmark the spatiotemporal structure of STWPs. In the course of our work over the past few years, we have settled on three measurement modalities that are necessary to definitively identify the salient characteristics of an STWP. These measurements acquire the following quantities: (1) the time averaged intensity $I(x,z)$; (2) the spatiotemporal spectral intensity $|\widetilde{\psi}(k_{x},\lambda)|^{2}$; and (3) the spatiotemporal intensity $I(x,z;t)$, from which the marginal spatial and temporal profiles can be extracted. Additionally, determining the STWP group-velocity $\widetilde{v}$ can be critical in some applications (see Section~\ref{sec:GroupVelocity}).

\subsubsection{Time-averaged intensity}

\begin{figure*}[t]
    \centering
    \includegraphics[width=160mm]{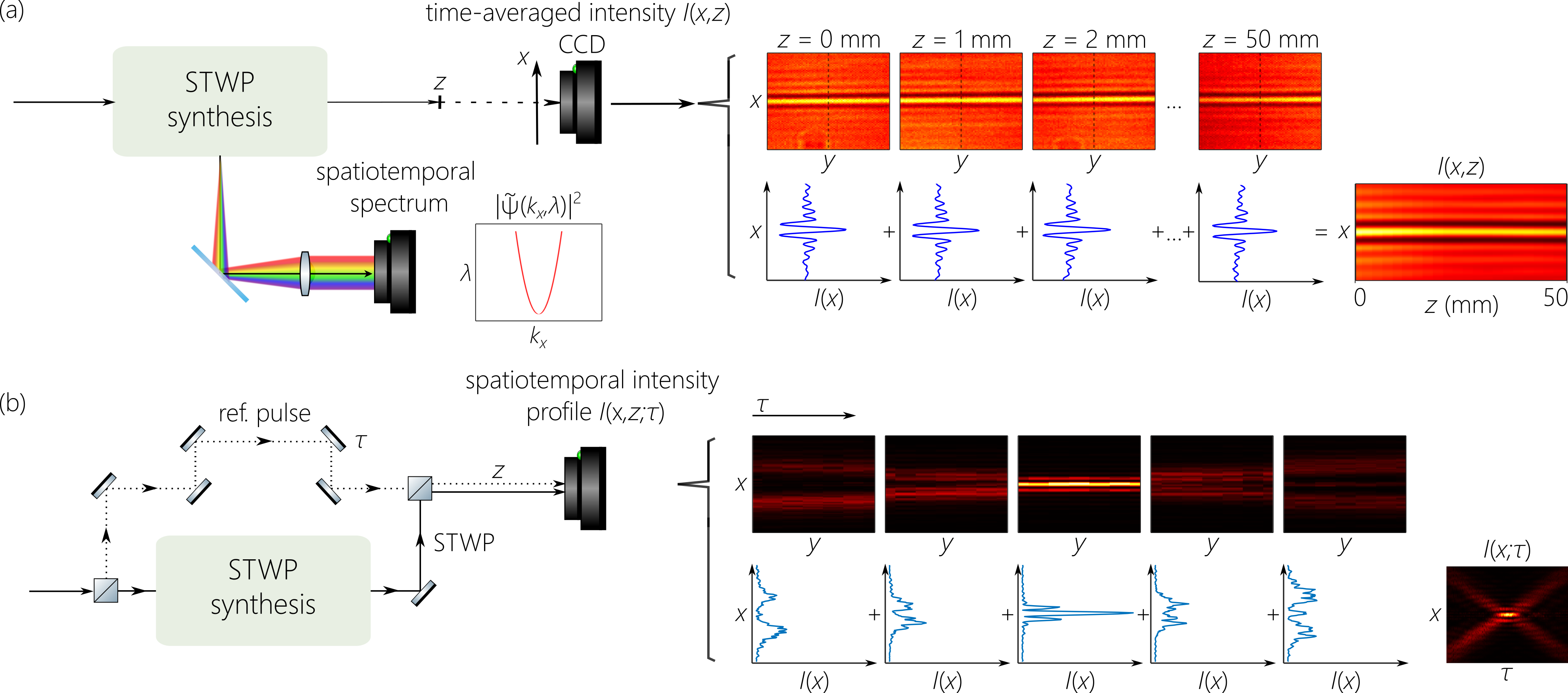}
    \caption{(a) Schematic of the measurement configuration for acquiring the axial evolution of the time-averaged intensity $I(x,z)$ and the spatiotemporal spectrum $|\widetilde{\psi}(k_{x},\lambda)|^{2}$. To acquire $I(x,z)$, the intensity profile $I(x,y,z)$ is recorded at each axial plane $z$, and a 1D profile $I(x,z)$ is extracted. To acquire $|\widetilde{\psi}(k_{x},\lambda)|^{2}$ a combination of a temporal Fourier transform (via a grating; not shown) and a spatial Fourier transform (via a lens; shown) is implemented to yield the spatiotemporal intensity profile at the CCD (this spectrum for an STWP takes the form of a parabola). (b) Schematic of the measurement configuration for reconstructing the spatiotemporal intensity profile $I(x,z;\tau)$ at a fixed axial plane $z$. At fixed $z$, the delay $\tau$ in the path of the reference pulse in the two-path interferometer is scanned. At each setting for $\tau$, the spatially resolved interference fringes resulting from the overlap of the STWP and the reference pulse are recorded. The spatiotemporal intensity profile $I(x,z;\tau)$ is then reconstructed from the visibility of these fringes.}
    \label{Fig:CharacterizationSchematics}
\end{figure*}

The first step in the characterization procedure is to measure the axial evolution of the time-averaged intensity $I(x,z)\!=\!\int\!dt|\psi(x,z;t)|^{2}$ by scanning a CCD camera along the propagation axis $z$. In the case of light sheets, the transverse intensity profile is uniform along $y$ but localized along $x$. One may thus sample the intensity profile along $y$ or integrate the intensity along $y$ to extract a 1D intensity profile along $x$ for each axial plane $z$ [Fig.~\ref{Fig:CharacterizationSchematics}(a)]. The spatial resolution of the CCD camera must be sufficient to capture the transverse spatial features in the STWP.

Three features are to be sought in the time-averaged intensity $I(x,z)$, which signify successful synthesis of an STWP. First, a hallmark of an STWP is that the intensity profile is independent of $z$ for an extended length $L_{\mathrm{max}}$. This length usually far exceeds the Rayleigh range associated with a conventional optical beam having the same spatial width $\Delta x$ of the central feature of the STWP \cite{Kondakci16OE,Kondakci17NP,Kondakci19OL,Yessenov22AOP}. Second, one must observe at each axial plane a transversely extended pedestal atop of which is a localized spatial feature of width $\Delta x$. Third, the width $\Delta x$ of this spatial feature should be approximately half the beam width at the pulse center which will extracted from the time-resolved measurement to be described below.

\subsubsection{Spatiotemporal spectrum}

Another crucial aspect of an STWP to be validated is its spatiotemporal spectrum $\widetilde{\psi}(k_{x},\Omega)$. Several features of interest are sought in this spectrum. (1) The measured spatiotemporal spectral intensity $|\widetilde{\psi}(k_{x},\Omega)|^{2}$ should take on a parabolic form. (2) This measurement reveals the temporal bandwidth $\Delta\omega$, and the spatial bandwidth $\Delta k_{x}$. (3) The \textit{sign} of the curvature of the spectral parabola distinguishes between the subluminal and superluminal regimes. (4) The \textit{magnitude} of the curvature of the parabolic spectrum indicates the spectral tilt angle $\theta$, and thus the group velocity $\widetilde{v}\!=\!c\tan{\theta}$. Measuring this spectrum requires taking both spatial and temporal Fourier transforms of the STWP. One may exploit the fact that the spectrum is already spatially resolved after traversing the SLM in the configuration shown in Fig.~\ref{Fig:ComparisonOfPulseShaperAndSTWP}(b). Therefore, one need implement only a spatial Fourier transform on that field to obtain $|\widetilde{\psi}(k_{x},\Omega)|^{2}$ or $|\widetilde{\psi}(k_{x},\lambda)|^{2}$ [Fig.~\ref{Fig:CharacterizationSchematics}(a)].  

\subsubsection{Spatiotemporal intensity profile}

Previous methodologies for reconstructing the spatiotemporal profile of propagation-invariant wave packets include spatially resolved ultrafast pulse characterization techniques \cite{Bowlan09OL,Lohmus12OL,Piksarv12OE} and self-reference interferometry \cite{Dallaire09OE,Kondakci17NP}. In our work here, to reconstruct the spatiotemporal intensity profile $I(x,z;\tau)$ at a fixed axial plane $z$, we embed the STWP synthesizer in a two-path (Mach Zehnder) interferometer as shown in Fig.~\ref{Fig:CharacterizationSchematics}(b). A portion of the initial generic pulsed beam is split to be used as a reference pulse and is directed through an optical delay line $\tau$, while the STWP synthesizer is placed in the second arm. The reference pulse and synthesized STWP are overlapped at a beam splitter and directed to a CCD camera. The delay $\tau$ is adjusted to overlap the two wave packets (reference and STWP) in space and time, thereby yielding spatially resolved fringes captured by the CCD camera.

The field at the detector $E(x,z;t,\tau)$ is a superposition of the STWP field $E_{\mathrm{ST}}(x,z;t)$ and reference-pulse field $E_{\mathrm{ref}}(x,z;t-\tau)$. Approximating the reference pulse as a plane wave spatially, we can drop its $x$-dependence. In this case, both STWP and reference pulse are propagation invariant, although they travel at different group velocities ($\widetilde{v}$ for the STWP, and $c$ for the reference pulse). We place the detector at the plane at which the two wave packets overlap, in which case we can drop their $z$-dependence, leading to:
\begin{equation}
E(x;t,\tau)=E_{\mathrm{ref}}(t-\tau)+E_{\mathrm{ST}}(x;t),
\end{equation}
where $E_{\mathrm{ref}}(t)\!=\!e^{-i\omega_{\mathrm{o}}t}\psi_{\mathrm{ref}}(t)$, $\psi_{\mathrm{ref}}(t)\!=\!\int\!d\Omega\,\widetilde{\psi}_{\mathrm{ref}}(\Omega)e^{-i\Omega t}$, $E_{\mathrm{ST}}(x;t)\!=\!e^{-i\omega_{\mathrm{o}}t}\psi_{\mathrm{ST}}(x;t)$, $\psi_{\mathrm{ST}}(x;t)\!=\!\int\!d\Omega\,\widetilde{\psi}(\Omega)e^{-i\{\Omega t-k_{x}(\Omega)x\}}$. The slow detector (a CCD camera here) records the time-averaged intensity, $I(x;\tau)\!=\!\int\!dt\,|E(x;t,\tau)|^{2}$, which can be written as:
\begin{equation}
I(x;\tau)=I_{\mathrm{ref}}+I_{\mathrm{ST}}(x)+2\mathrm{Re}\{e^{-i\omega_{\mathrm{o}}t}R(x;\tau)\},
\end{equation}
where $I_{\mathrm{ref}}\!=\!\int\!dt\,|\psi_{\mathrm{ref}}(t)|^{2}\!=\!\int\!d\Omega\,|\widetilde{\psi}_{\mathrm{ref}}(\Omega)|^{2}$, $I_{\mathrm{ST}}(x)\!=\!\int\!dt\,|\psi_{\mathrm{ST}}(x;t)|^{2}$, and $R(x;\tau)=|R(x;\tau)|e^{i\phi(x;\tau)}\!=\!\int\!dt\,\psi^{*}_{\mathrm{ref}}(t-\tau)\psi_{\mathrm{ST}}(x;t)$.

The spatially resolved interference pattern observable in $I(x;\tau)$ is given by:
\begin{equation}
I(x,\tau)\propto1+V(x;\tau)\cos[\omega_{\mathrm{o}}\tau+\phi(x;\tau)],
\end{equation}
where the spatially varying visibility $V$ at any given delay $\tau$ is given by:
\begin{equation}
V(x;\tau)=\frac{2|R(x;\tau)|}{I_{\mathrm{ref}}+I_{\mathrm{ST}}(x)}\approx\frac{2\sqrt{I_{\mathrm{ref}}}}{I_{\mathrm{ref}}+I_{\mathrm{ST}}(x)}|\psi_{\mathrm{ST}}(x;\tau)|.
\end{equation}
The approximation used is the assumption that the temporal bandwidth of the STWP is smaller than that of the reference pulse (or the reference pulse is temporally narrower than the STWP to allow sampling it interferometrically), which is typically the case in our experiments. Furthermore, when $I_{\mathrm{ref}}\!\gg\!I_{\mathrm{ST}}(x)$ at any $x$, then we have simply $V(x;\tau)\!\propto\!|\psi_{\mathrm{ST}}(x;\tau)|$. That is, mapping the visibility of the spatially resolved interference pattern while scanning $\tau$ resulting from the superposition of the STWP and the reference pulse unveils the magnitude of the STWP envelope, and thus allows us to reconstruct its spatiotemporal intensity profile.

\section{Experimental arrangement}

In this Section we describe in detail the experimental apparatus to synthesize and characterize STWPs according to the conceptual schemes outlined above.

\subsection{Beam synthesis}

\begin{figure*}[t!]
    \centering
    \includegraphics[width=160mm]{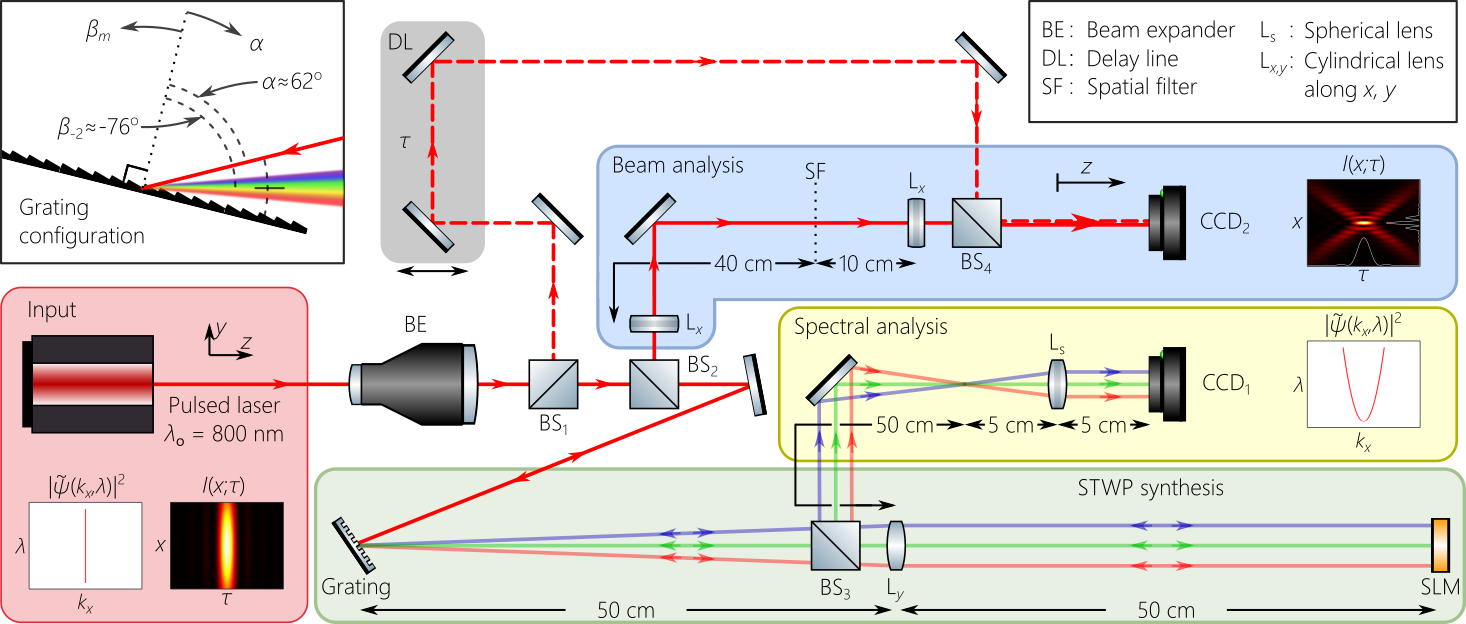}
    \caption{Schematic of the setup design for synthesizing and characterizing STWPs from a pulsed plane wave. Distinct functionalities in the setup are bounded and highlighted with different-color boxes, including Input, STWP synthesis, Spectral analysis, and Beam analysis. The top left inset shows the angular settings for the optical beams at the diffraction grating (Grating configuration).}
    \label{Fig:Setup}
\end{figure*}

As shown in Fig.~\ref{Fig:Setup}, we start with pulses from a mode-locked Ti:sapphire femtosecond laser (Tsunami; Spectra Physics): central wavelength is $\lambda_{\mathrm{o}}\!\approx\!800$~nm, bandwidth $\Delta\lambda\!\approx\!10$~nm (FWHM), and pulse width $\Delta T\!\approx\!100$~fs. The laser beam is directed to a reflective diffraction grating (1200~lines/mm, $25\times25$~mm$^{2}$-area). To secure high spectral resolution from the grating, we first increase the transverse spatial extent of the beam via a beam expander (Thorlabs BE02-05B) to match the grating width of 25~mm. A portion of the expanded beam is split off by beam splitter BS$_{1}$ to serve as the reference pulse for the spatiotemporal intensity profile reconstruction. This reference pulse is directed to a motorized optical delay line $\tau$ after which it is combined with the synthesized STWP at beam splitter BS$_{4}$.

The spatially expanded input pulse exiting the other output port from BS$_{1}$ is directed to the `STWP synthesis' arrangement. Beam splitter BS$_{2}$ separates the incident plane-wave pulse from the returned STWP. The incident plane-wave pulse is directed to a reflective grating at an incident angle $\alpha\!=\!62^{\circ}$ from its normal. We select the $m\!=\!-2$ diffraction order at an angle $\beta_{2}\!\approx\!-76^{\circ}$ from the normal, thus resulting in an angle $\approx\!14^{\circ}$ between the incident and diffracted beams [Fig.~\ref{Fig:Setup}, Grating configuration]. The diffracted wavelengths are angularly spread along the $y$-axis, and the spatially resolved spectrum is focused along $y$ via a cylindrical lens L$_{y}$ of focal length $f\!=\!50$~cm in a $2f$ configuration. A reflective, phase-only SLM (Hamamatsu, X10468) is placed at the focal plane of this cylindrical lens [Fig.~\ref{Fig:Setup}, STWP synthesis]. The SLM imparts a 2D phase distribution $\Phi(x,y)$ to the spectrally resolved wave front, and the phase-modulated field is retro-reflected from the SLM through the cylindrical lens back to the grating where the wavelengths are recombined and the pulse is reconstituted, thereby yielding the STWP in the form of a light sheet.

Although the temporal bandwidth of the initial laser pulses is $\Delta\lambda\!\sim\!10$~nm, the bandwidth of the synthesized STWPs is narrower. This is because the spatially resolved spectrum may spread beyond the width of the SLM active area. Consequently, only a portion of the spectrum is utilized in the synthesis procedure and retro-reflected by the SLM. In our experiments, we have an STWP bandwidth of $\Delta\lambda\!\approx\!2$~nm (extending over $798-800$~nm). Because the truncated STWP bandwidth is narrower than that of the initial laser pulses, the temporal spectral intensity is roughly flat over the reduced bandwidth selected. 

Although the spatiotemporal pulse shaper in Fig.~\ref{Fig:Setup} operates in a reflective configuration, it may of course be alternatively `unfolded' to operate in a transmissive modality, using either a transmissive SLM or a phase plate \cite{Kondakci18OE,Bhaduri18OL}. This is particularly useful in those cases where SLMs are not available (e.g., in the mid-infrared \cite{Yessenov20OSAC}) or not useful (e.g., at high power levels or when high spatial resolution is needed).

\subsection{Phase modulation by the SLM}

The design of the 2D phase pattern $\Phi(x,y)$ imparted to the spectrally resolved wave front is the crucial step in the synthesis procedure. In our case the SLM has $N_{y}\times N_{x}\!=\!800\times600$~pixels over an area $16\times12$~mm$^{2}$ (the spatially resolved wavelengths $\lambda(y)$ are arranged linearly along the $y$-axis). We index the wavelength $\lambda$ along $y$ as:
\begin{equation}\label{eq:DiscreteWavelength}
\lambda(i)=\lambda_\mathrm{o}\pm\frac{i-1}{N_y-1}\Delta\lambda,
\end{equation}
where $1\!\leq\!i\!\leq\!N_{y}$, $\Delta\lambda$ is the temporal bandwidth intercepted by the SLM from the incident spectrally resolved field ($\Delta\lambda\!\approx\!2$~nm here). The choice of the positive or negative sign in Eq.~\ref{eq:DiscreteWavelength} depends on whether the subluminal or superluminal regime for the STWP group velocity is targeted, respectively. In all cases, the spectrum incident on the SLM extends from $\lambda\!=\!798$~nm to $\lambda\!=\!800$~nm. For subluminal STWPs ($\widetilde{v}\!<\!c$ and $0\!<\!\theta\!<\!45^{\circ}$), we select the positive sign in Eq.~\ref{eq:DiscreteWavelength}, $\lambda(i)\!=\!\lambda_\mathrm{o}+\tfrac{i-1}{N_y-1}\Delta\lambda$, so that $\lambda(1)\!=\!\lambda_{\mathrm{o}}\!=\!798$~nm, $\lambda(i)\!\geq\!\lambda_{\mathrm{o}}$, and $\lambda(N_{y})\!=\!\lambda_{\mathrm{o}}+\Delta\lambda\!=\!800$~nm. In other words, $\lambda_{\mathrm{o}}$ corresponds to the shortest wavelength in the temporal spectrum of a subluminal STWP. For superluminal STWPs ($\widetilde{v}\!>\!c$ or $\widetilde{v}\!<\!0$, corresponding to $45^{\circ}\!<\!\theta\!<\!180^{\circ}$), we select the negative sign in Eq.~\ref{eq:DiscreteWavelength}, $\lambda(i)\!=\!\lambda_\mathrm{o}-\tfrac{i-1}{N_y-1}\Delta\lambda$, so that $\lambda(1)\!=\!\lambda_{\mathrm{o}}-\Delta\lambda\!=\!798$~nm, $\lambda(i)\!\leq\!\lambda_{\mathrm{o}}$, and $\lambda(N_{y})\!=\!\lambda_{\mathrm{o}}\!=\!800$~nm. In other words, $\lambda_{\mathrm{o}}$ corresponds to the longest wavelength in the temporal spectrum of a subluminal STWP. In both cases of subluminal and superluminal STWPs, the spatial frequency $k_{x}\!=\!0$ is associated with $\lambda_{\mathrm{o}}$.

Each column along the $x$-axis is divided into two halves around $x\!=\!0$. The upper half of the SLM has $800\times300$~pixels and is used to impart a phase pattern of the form $\Phi(x, y)\!=\!k_{x}(\lambda(y))x$ corresponding to positive spatial frequencies, and the lower half a phase $\Phi(x,y)\!=\!-k_{x}(\lambda(y))x$ corresponding to negative spatial frequencies, where $k_{x}(\lambda)\!=\!\tfrac{2\pi}{\lambda_{\mathrm{o}}}\sqrt{2(1-\cot{\theta})(\tfrac{\lambda_{\mathrm{o}}}{\lambda}-1)}$. We write the vertical position as $x(j)\!=\!(j-\tfrac{N_{x}}{2})d$, where $d$ is the pixel pitch; here $1\!\leq\!j\!\leq\!\tfrac{N_{x}}{2}$ corresponds to negative-$x$, whereas $\tfrac{N_{x}}{2}+1\!\leq\!j\!\leq\!N_{x}$ corresponds to positive-$x$. The SLM phase $\Phi(i,j)\!=\!Mk_{x}(i)x(j)$ can thus be written explicitly as follows:
\begin{equation}\label{eq:PhasePattern}
\Phi(i,j)=
2\pi M\frac{d}{\lambda_\mathrm{o}}\left(j-\frac{N_{x}}{2}\right)\sqrt{2(1-\cot{\theta})\left(\frac{\lambda_\mathrm{o}}{\lambda(i)}-1\right)};
\end{equation}
we discuss the significance of the additional factor $M$ below.

\begin{figure*}[t!]
    \centering
    \includegraphics[width=157mm]{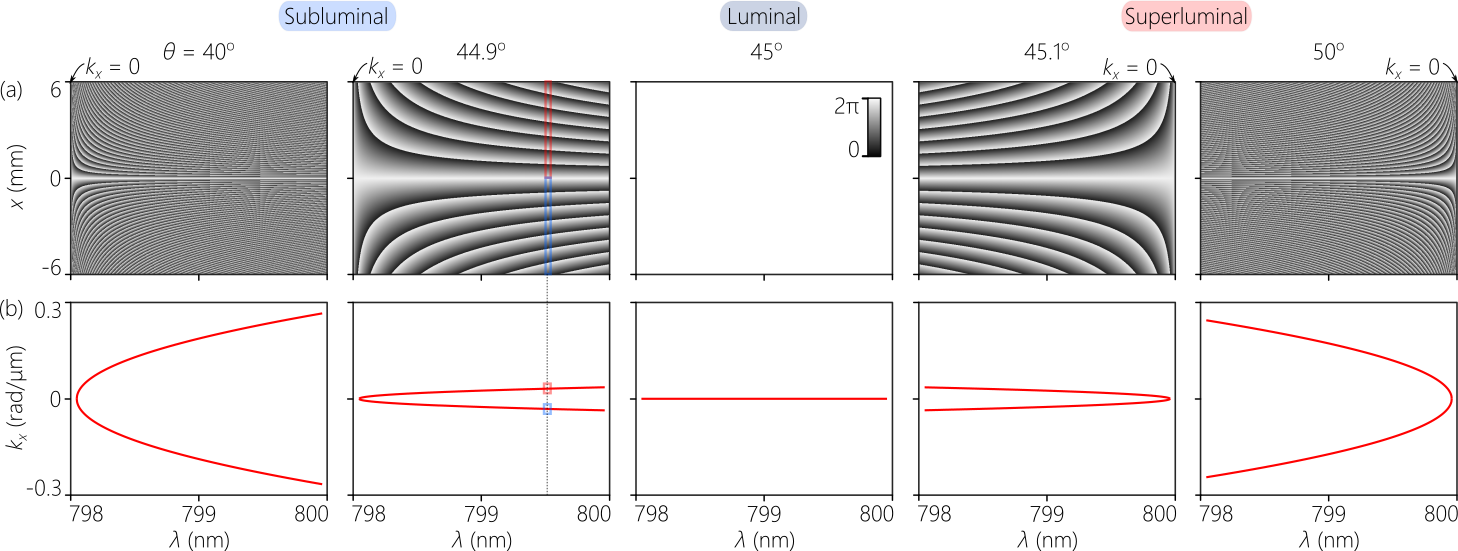}
    \caption{(a) SLM phase distributions $\Phi$ corresponding to subluminal ($\theta\!=\!40^{\circ}$ and $44.9^{\circ}$) and superluminal ($\theta\!=\!45.1^{\circ}$ and $50^{\circ}$) STWPs, taking into account demagnification by a factor of $M\!=\!\frac{1}{4}$. The case of $\theta\!=\!45^{\circ}$ corresponds to a plane-wave pulse, whereupon $\Phi\!=\!0$. We identify the $k_{x}\!=\!0$ component in each panel. (b) Spatiotemporal spectra corresponding to the phase distributions in (a). In the panel for $\theta\!=\!44.9^{\circ}$, we highlight in the phase distribution a positive and negative spatial frequency pair $\pm k_{x}$ (red and blue rectangles, respectively), and show the corresponding spatial frequencies in the spatiotemporal spectrum (identified by the corresponding small red and blue boxes).}
    \label{Fig:SLMPhasePatterns}
\end{figure*}

We plot in Fig.~\ref{Fig:SLMPhasePatterns}(a) examples of the phase distribution $\Phi$ imparted by the SLM. We note several general features about these phase distributions. First, the overall structure of the phase distribution along the horizontal $\lambda$-axis flips when switching from the subluminal ($\theta\!<\!45^{\circ}$) to the superluminal ($\theta\!>\!45^{\circ}$) regimes. Therefore, whereas the constant-phase $k_{x}\!=\!0$ component is associated with the shortest wavelength (highest temporal frequency) in the spectrum in the subluminal regime, it is associated with the longest wavelength (lowest temporal frequency) in the superluminal regime; see Fig.~\ref{Fig:LightCones}(b). Second, as the spectral tilt angle $\theta$ deviates away from the luminal condition $\theta\!=\!45^{\circ}$, larger values of $k_{x}$ are required to construct the STWP, and thus more rapid variations occur in the SLM phase distributions. For example, compare $\Phi(x,y)$ for $\theta\!=\!45.1^{\circ}$ that contains only slow spatial variations in the phase distribution to the case $\theta\!=\!50^{\circ}$ that features rapid spatial variations. At $\theta\!=\!45^{\circ}$, the phase distribution is set to $\Phi\!=\!0$, and the resulting STWP is a plane-wave pulse with $\Delta\lambda\!\approx\!2$~nm; no spatiotemporal structure has been introduced here into the field. The limited numerical aperture of the SLM sets a maximum value of $k_{x}$ that can be realized directly. This in turn sets limits on the maximum deviation of the spectral tilt angle $\theta$ from $45^{\circ}$, and therefore determines the smallest spatial feature size in the synthesized STWP. One may further expand the reach of this approach beyond the limit set by the SLM numerical aperture, and thus introduce a narrower spatial peak for the STWP, by adding a demagnifying telescope system with demagnification factor $M$ (Fig.~\ref{Fig:Setup}). This is the additional factor in Eq.~\ref{eq:PhasePattern}. 

\subsection{Spatiotemporal spectral analysis}

We exploit the fact that the spectrum of the field retro-reflected from the SLM is already spatially resolved, and split off a portion of this field at beam splitter BS$_{3}$ to measure the spatiotemporal spectrum $|\widetilde{\psi}(k_{x},\lambda)|^{2}$ at CCD$_{1}$ (Imaging Source DMK 33UX178). Considering the field distribution along the $y$-axis (where the wavelengths are spread), the field experiences a sequence of two lenses in its path between SLM and CCD$_{1}$: cylindrical lens L$_{y}$ with $f\!=\!50$~cm and spherical lens L$_{\mathrm{s}}$ with $f\!=\!5$~cm, arranged in a $4f$ imaging configuration, so that the focused, horizontally spread temporal spectrum at the SLM plane is mapped to CCD$_{1}$ except for a change in scale [Fig.~\ref{Fig:Setup}; Spectral analysis]. Considering the field distribution along the $x$-axis, the field experiences only a single lens L$_{\mathrm{s}}$ in a $2f$ Fourier-transforming configuration. As such, CCD$_{1}$ registers the spatial Fourier transform of the field along $x$ after acquiring the phase distribution $\Phi$ from the SLM. Each wavelength $\lambda$ is associated with a pair of spatial frequencies $\pm k_{x}(\lambda)$, so that the spatial Fourier transform for each wavelength comprises two points located at $x\!=\!\pm\tfrac{k_{x}}{2\pi}\lambda f$, where $f$ is the focal length of L$_{\mathrm{s}}$. This procedure yields the spatiotemporal spectrum $|\widetilde{\psi}(k_{x},\lambda)|^{2}$, which takes the form of a parabola if the correct phase distribution $\Phi$ has been implemented at the SLM. 

We plot in Fig.~\ref{Fig:SLMPhasePatterns}(b) the spatiotemporal spectra $|\widetilde{\psi}(k_{x},\lambda)|^{2}$ associated with the phase distributions in Fig.~\ref{Fig:SLMPhasePatterns}(a). These are the spectra that would be captured with the above-described spectral analysis system. Note that two points are associated with each wavelength $\lambda$, corresponding to $\pm k_{x}(\lambda)$, except at the degenerate point $\lambda\!=\!\lambda_{\mathrm{o}}$ where $k_{x}(\lambda_{\mathrm{o}})\!=\!0$. The orientation of the parabola flips along the $\lambda$-axis when switching from the subluminal to superluminal regimes, and the parabola curvature decreases away from the luminal condition $\theta\!=\!45^{\circ}$. At $\theta\!=\!45^{\circ}$, the spatiotemporal spectrum is that of a plane-wave pulse ($\Phi\!=\!0$).

\subsection{Beam analysis}

After recombining the wavelengths at the grating and reconstituting the pulse, the STWP is formed and reflected at BS$_{2}$ [Fig.~\ref{Fig:Setup}, Beam analysis]. A two-lens telescope system adjusts the transverse size of the STWP along $x$ in physical space, after which it travels diffraction-free along the $z$-axis. The focal length of the first cylindrical lens is $f\!=\!40$~cm and that of the second is $f\!=\!10$~cm, thereby yielding $M\!=\!\tfrac{1}{4}$. Moreover, this telescope system provides the opportunity to spatially filter the zeroth diffraction order from the SLM at the Fourier plane between the two lenses. We place a thin wire along the center of the Fourier plane [SF in Fig.~\ref{Fig:Setup}] to block the $k_{x}\!=\!0$ component remaining in the beam (this component is a consequence of the finite diffraction efficiency of the SLM). Of course, this spatial filter will also block spatial frequencies in the vicinity of $k_{x}\!=\!0$. 

We capture the time-averaged transverse intensity profile $I(x,y,z)$ at each axial plane $z$ via a CCD camera (Imaging Source DMK 27BUP031), while blocking the path of the delay line. The CCD is mounted on a motorized linear stage (Thorlabs LTS150) that allows scanning the CCD along the optical axis $z$. From these captured intensity profiles, we construct the axial evolution of the STWP intensity $I(x,z)$ as illustrated in Fig.~\ref{Fig:CharacterizationSchematics}(a).

\subsection{Spatiotemporal intensity reconstruction}

The reference pulse split off at BS$_{1}$ after the beam expander [Fig.~\ref{Fig:Setup}] is directed to a motorized delay line (Thorlabs DDS300) to introduce an optical delay $\tau$. This reference pulse is combined with the synthesized STWP at BS$_{4}$, after which the two wave packets travel together to CCD$_{2}$. Effectively, we have a Mach Zehnder interferometer between BS$_{1}$ and BS$_{4}$, corresponding to that illustrated in Fig.~\ref{Fig:CharacterizationSchematics}(b). Along the reference path, the plane-wave pulse traverses the delay line $\tau$, and along the second path it enters the STWP synthesizer followed by the two-lens telescope.

At this point, the STWP has narrower temporal spectrum ($\Delta\lambda\!\approx\!2$~nm, longer pulse width) than the reference pulse ($\Delta\lambda\!\approx\!10$~nm) because of the spectral filtering occurring at the SLM. Furthermore, the spatial features in the STWP are much narrower than the spatial width of the reference beam. By scanning the relative delay $\tau$, we reconstruct the spatiotemporal intensity profile of the STWP $I(x,z;\tau)$ at any fixed axial plane $z$ from the measured visibility of the spatially resolved fringes resulting from the interference of the two fields, as illustrated conceptually in Fig.~\ref{Fig:CharacterizationSchematics}(b). To improve the signal-to-noise ratio, at each delay $\tau$ we average the results over several samples taken in the vicinity of $\tau$. 

\section{Spatiotemporal spectrum: Model and measurements}

\begin{figure*}[t!]
    \centering
    \includegraphics[width=157mm]{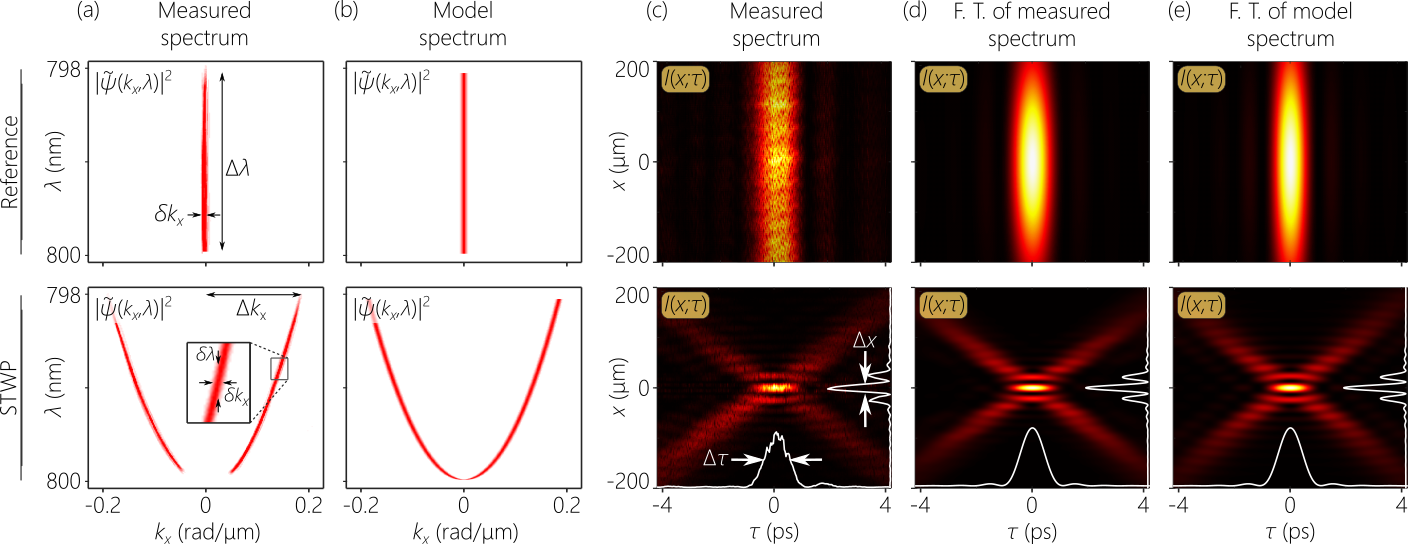}
    \caption{Spatiotemporal spectra and intensity profiles for the reference pulse (top row) and an STWP (second row); $\Delta\lambda\!\approx\!2$~nm, carrier wavelength $\lambda_{\mathrm{o}}\!\approx\!800$~nm, and $\theta\!=\!50^{\circ}$ for the STWP. (a) Measured spatiotemporal spectrum and (b) the model spatiotemporal spectrum; here $\delta k_{x}\!\approx\!4$~m$\mathrm{rad/\mu m}$ for the reference pulse. The inset highlights the spectral uncertainty. (c-e) Spatiotemporal intensity profiles: (c) experimentally reconstructed from interferometric measurements; (d) obtained by taking the Fourier transform of the measured spatiotemporal spectra in (a); and (e) by taking the Fourier transform of the model spatiotemporal spectra in (b).}
    \label{Fig:SpatioTemporalSpectrumAndProfile}
\end{figure*}

We plot in Fig.~\ref{Fig:SpatioTemporalSpectrumAndProfile}(a) the measurement results for two wave packets: (1) a reference pulse (that differs from the reference pulse used in the interferometric detection scheme in its reduced bandwidth), which is approximately a plane-wave pulse, thus corresponding to $\theta\!=\!45^{\circ}$, and is obtained by idling the SLM, $\Phi(x,y)\!=\!0$; and (2) a superluminal STWP corresponding to $\theta\!=\!50^{\circ}$ obtained by implementing the SLM phase distribution shown in Fig.~\ref{Fig:SLMPhasePatterns}(a). Both wave packets have the same temporal bandwidth $\Delta\lambda\!\approx\!2$~nm, but have different spatial bandwidths and distinct spatiotemporal structures.

The spatiotemporal spectrum $\widetilde{\psi}(k_{x},\lambda)$ for the reference pulse is a 2D distribution that is separable with respect to its spatial and temporal DoFs $\widetilde{\psi}(k_{x},\lambda)\!\rightarrow\!\widetilde{\psi}_{t}(\lambda)\delta(k_{x})$. For an ideal STWP, this spectrum becomes a 1D distribution of the form $\widetilde{\psi}(k_{x},\lambda)\!\rightarrow\!\widetilde{\psi}_{t}(\lambda)\delta(k_{x}-k_{x}(\lambda))$, where $k_{x}(\lambda)$ is the spatial frequency associated with each wavelength in the STWP according to Eq.~\ref{Eq:Parabola}. To compare the measurement results to theoretical expectations, we construct spatiotemporal spectral models for the reference pulse and the STWP [Fig.~\ref{Fig:SpatioTemporalSpectrumAndProfile}(b)], which we then compare to the measured spectra.

\subsection{Reference pulse}

For the reference pulse, the separable spatiotemporal spectrum in the ideal limit is $\widetilde{\psi}(k_{x},\lambda)\!\approx\!\widetilde{\psi}_{t}(\lambda)\delta(k_{x})$; $\widetilde{\psi}_{t}(\lambda)$ is the temporal spectrum with a bandwidth $\Delta\lambda\!\approx\!2$~nm, corresponding to a pulsewidth $\Delta T\!\approx\!1.4$~ps. Because the temporal bandwidth is filtered out from the broader spectrum of the original laser pulses by the SLM aperture, an adequate model of the spectrum is given by a flat top function: $\widetilde{\psi}_{t}(\lambda)\!\propto\!\mathrm{rect}(\tfrac{\lambda-\lambda_{\mathrm{c}}}{\Delta\lambda})$, where $\mathrm{rect}(\tfrac{x}{W})$ is a flat distribution of width $W$ centered at $x\!=\!0$, and $\lambda_{\mathrm{c}}$ is the central wavelength of the pulse spectrum. In practice, the finite system apertures (25~mm here) result in a finite (but narrow) spatial bandwidth $\delta k_{x}$, so that $\widetilde{\psi}(k_{x},\lambda)\!\approx\!\widetilde{\psi}_{t}(\lambda)\widetilde{\psi}_{x}(k_{x})$. We model the spatial spectrum as a Gaussian function, leading to a spatiotemporal spectrum of the form
\begin{equation}
\widetilde{\psi}(k_{x},\lambda)=\widetilde{\psi}_{x}(k_{x})\widetilde{\psi}_{t}(\lambda)\propto\mathrm{rect}\left(\frac{\lambda-\lambda_{\mathrm{c}}}{\Delta\lambda}\right)\exp\left\{\frac{-k_{x}^{2}}{2(\delta k_{x})^{2}}\right\}.    
\end{equation}
In the first row of Fig.~\ref{Fig:SpatioTemporalSpectrumAndProfile}(a,b), we plot the measured and modeled spatiotemporal spectra of the plane-wave pulse and observe excellent agreement between the two.  

\subsection{STWP}

We plot in the second row of Fig.~\ref{Fig:SpatioTemporalSpectrumAndProfile} the corresponding spatiotemporal spectrum for the STWP. The STWP spatiotemporal spectrum is no longer separable with respect to the spatial and temporal DoFs, and is ideally given by 
\begin{equation}
\widetilde{\psi}(k_{x},\lambda)=\widetilde{\psi}_{t}(\lambda)\delta(k_{x}-k_{x}(\lambda))\propto\mathrm{rect}\left(\frac{\lambda-\lambda_{\mathrm{c}}}{\Delta\lambda}\right)\delta(k_{x}-k_{x}(\lambda)),    
\end{equation}
where $k_{x}(\lambda)$ is given by Eq.~\ref{Eq:Parabola}. The result is a parabola-shaped spatiotemporal spectrum, and the measured spectrum plotted in Fig.\ref{Fig:SpatioTemporalSpectrumAndProfile}(a) is in excellent agreement with the theoretical model plotted in Fig.~\ref{Fig:SpatioTemporalSpectrumAndProfile}(b). Note, however, that the measured spectrum does not correspond to an ideal delta-function correlation between $k_{x}$ and $\lambda$. Indeed, such an ideal spectrum is not realizable in practice because it implies an infinite energy \cite{Sezginer85JAP}. Rather, there is a finite `spectral uncertainty': an intrinsic `fuzziness' in the association between the spatial and temporal frequencies. Each spatial frequency $k_{x}$ is \textit{not} strictly associated with a single wavelength $\lambda(k_{x})$, but is instead associated with a finite but narrow spectral uncertainty $\delta\lambda$ centered at the ideal wavelength $\lambda(k_{x})$. Alternatively, each wavelength $\lambda$ is not associated with a single spatial frequency $k_{x}(\lambda)$, rather it is associated with a finite spatial bandwidth $\delta k_{x}$ centered at the ideal spatial frequency $k_{x}(\lambda)$; see Fig.~\ref{Fig:SpatioTemporalSpectrumAndProfile}(a), inset. To accommodate this finite spectral uncertainty, we modify our spectral model, $\widetilde{\psi}(k_{x},\lambda)\!=\!\widetilde{\psi}_{t}(\lambda)\widetilde{g}(\lambda-\lambda(k_{x}))$, where $\widetilde{g}(\lambda)$ is a narrow spectral function of width $\delta\lambda$ (taken here to have a Gaussian profile).

\section{Spatiotemporal intensity model and measurements}

We plot in Fig.~\ref{Fig:SpatioTemporalSpectrumAndProfile}(c-e) the measured spatiotemporal intensity profiles $I(x,z;\tau)\!=\!|\psi(x,z;\tau)|^{2}$ for the reference pulse and the STWP, along with the theoretical expectations based on the above-described spatiotemporal spectral model. Because there is minimal change in the profile of either wave packet over an extended distance, we confine the measurements and theoretical predictions to a single axial plane $z\!=\!0$.

\subsection{Reference pulse}

The spatiotemporal intensity profile at $z\!=\!0$, $I(x;\tau)\!=\!I(x,z\!=\!0;\tau)$, is also separable with respect to the spatial and temporal DoFs for the reference pulse: $\psi(x,0;t)\!=\!\psi_{x}(x)\psi_{t}(t)$, where $\psi_{x}(x)$ and $\psi_{t}(t)$ are the spatial and temporal Fourier transforms of  $\widetilde{\psi}_{x}(k_{x})$ and $\widetilde{\psi}_{t}(\lambda)$, respectively. We plot in Fig.~\ref{Fig:SpatioTemporalSpectrumAndProfile}(c) the experimentally reconstructed spatiotemporal intensity profile $I(x;\tau)\!=\!|\psi(x,0;\tau)|^{2}$ obtained using the setup in Fig.~\ref{Fig:Setup} following the interferometric approach illustrated in Fig.~\ref{Fig:CharacterizationSchematics}(b). Next, we plot in Fig.~\ref{Fig:SpatioTemporalSpectrumAndProfile}(d) the spatiotemporal profile $I(x;\tau)$ calculated by taking the Fourier transform of the measured spatiotemporal spectrum plotted in Fig.~\ref{Fig:SpatioTemporalSpectrumAndProfile}(a). The agreement between this calculated spatiotemporal profile and the measured counterpart in Fig.~\ref{Fig:SpatioTemporalSpectrumAndProfile}(c) indicates that there is no phase structure or chirp \cite{Kondakci18PRL,Wong21} in the spatiotemporal structure in Fig.~\ref{Fig:SpatioTemporalSpectrumAndProfile}(a). Finally, in Fig.~\ref{Fig:SpatioTemporalSpectrumAndProfile}(e) we plot $I(x;\tau)$ calculated by taking the Fourier transform of the model spectrum shown in Fig.~\ref{Fig:SpatioTemporalSpectrumAndProfile}(b). With the model used above, this envelope takes the form $I(x;\tau)\!=\!|\psi(x,0;\tau)|^{2}$, where $\psi(x,0;\tau)\!\propto\!\mathrm{sinc}(\tfrac{\Delta\omega\tau}{2\pi})\exp\{-\tfrac{1}{2}x^{2}(\delta k_{x})^{2}\}$, and we define $\mathrm{sinc}(x)\!=\!\tfrac{\sin{\pi x}}{\pi x}$. The agreement between this profile and that in Fig.~\ref{Fig:SpatioTemporalSpectrumAndProfile}(c) confirms the appropriateness of the spatiotemporal spectral model proposed.

\subsection{STWP}

The spatiotemporal intensity profile for the STWP differs fundamentally from that of the reference pulse. The experimentally reconstructed profile of the STWP plotted in Fig.~\ref{Fig:SpatioTemporalSpectrumAndProfile}(c) is non-separable with respect to the spatial and temporal DoFs, and takes on the expected X-shaped structure common to all propagation-invariant wave packets in free space \cite{FigueroaBook14} (with the exception of those whose spectra feature structured phase modulation \cite{Kondakci18PRL,Wong21}, or in the presence of anomalous GVD \cite{Malaguti08OL,Hall23NPhys,Hall24XtoO}). The measured profile [Fig.~\ref{Fig:SpatioTemporalSpectrumAndProfile}(c)] is in excellent agreement with that obtained by taking the Fourier transform of the measured spatiotemporal spectrum [Fig.~\ref{Fig:SpatioTemporalSpectrumAndProfile}(d)], and that resulting from the Fourier transform of the spectral model [Fig.~\ref{Fig:SpatioTemporalSpectrumAndProfile}(e)]. We did not obtain a closed-form expression for the STWP spatiotemporal intensity profile profile. However, if the spectrum is taken to be a Gaussian function, a closed-form expression can be readily reached \cite{Yessenov19OE,Kondakci19OL}.

\section{Marginal spatial and temporal profiles}

Despite the differences between the structures of the spatiotemporal profiles of the reference pulse and the STWP [Fig.~\ref{Fig:SpatioTemporalSpectrumAndProfile}], some similarities nevertheless appear in the marginal profiles. Moreover, examining the marginal profiles provides further opportunity to confirm the validity of the theoretical model used for the STWPs. We investigate here the following marginal spatial and temporal distributions: the pulse profile at the beam center $I_{t}(\tau)\!=\!I(0,0;\tau)\!=\!|\psi_{t}(\tau)|^{2}$, where $\psi_{t}(\tau)\!=\!\int\!d\Omega\,\widetilde{\psi}_{t}(\Omega)e^{-i\Omega \tau}$ and $\widetilde{\psi}_{t}(\Omega)\!=\!\int\!dk_{x}\,\widetilde{\psi}(k_{x},\Omega)$; and the beam profile at the pulse center $I_{x}(x)\!=\!I(x,0;0)\!=\!|\psi_{x}(x)|^{2}$, where $\psi_{x}(x)\!=\!\int\!dk_{x}\,\widetilde{\psi}_{x}(k_{x})e^{ik_{x}x}$, and $\widetilde{\psi}_{x}(k_{x})\!=\!\int\!d\Omega\,\widetilde{\psi}(k_{x},\Omega)$. We examine the time-averaged intensity $I(x,z)\!=\!\int\!d\tau\;I(x,z;\tau)$ in the next Section. Recall that the results in Fig.~\ref{Fig:SpatioTemporalSpectrumAndProfile} confirmed that there is no spectral phase structure in the spatiotemporal spectra for either the reference pulse or the STWP. 

\begin{figure*}[t]
    \centering
    \includegraphics[width=157mm]{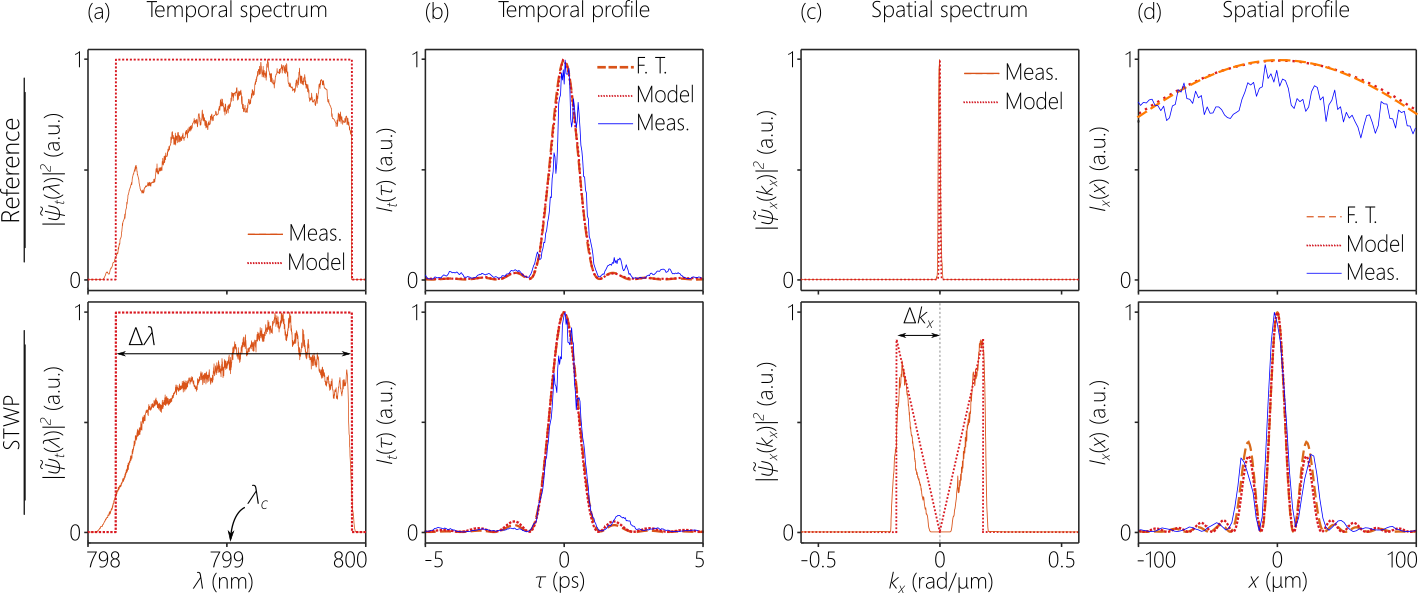}
    \caption{Marginal temporal and spatial spectra, and marginal profiles in space and time for the reference pulse (first row) and for an STWP (second row), corresponding to those in Fig.~\ref{Fig:SpatioTemporalSpectrumAndProfile}. (a) Marginal temporal spectrum $|\widetilde{\psi}_{t}(\lambda)|^{2}$ obtained from the measured spatiotemporal spectrum (`Meas.'), compared to the spectral model (`Model'). (b) Marginal temporal profile $I_{t}(\tau)$ obtained from the measured spatiotemporal profile in Fig.~\ref{Fig:SpatioTemporalSpectrumAndProfile}(c) (`Meas.'); by taking the Fourier transform of the measured marginal temporal spectrum plotted in (a) (`F.T.'); and by taking the Fourier transform of the spectral model in (a) (`Model'). (c) Marginal spatial spectrum $|\widetilde{\psi}(k_{x})|^{2}$ obtained from the measured spatiotemporal spectrum in Fig.~\ref{Fig:SpatioTemporalSpectrumAndProfile}(a) (`Meas.'), compared to the spectral model (`Model'). (d) Marginal spatial profile $I_{x}(x)$ obtained from the measured spatiotemporal profile in Fig.~\ref{Fig:SpatioTemporalSpectrumAndProfile}(c) (`Meas.'); by taking the Fourier transform of the measured marginal spatial spectrum plotted in (c) (`F.T.'); and by taking the Fourier transform of the spectral model in (c) (`Model').}
    \label{Fig:SeparateSpatialAndTemporalProfiles}
\end{figure*}

\subsection{Marginal temporal profile: Pulse profile at the beam center}

\subsubsection{Reference pulse}

We first plot $|\widetilde{\psi}_{t}(\lambda)|^{2}$ in Fig.~\ref{Fig:SeparateSpatialAndTemporalProfiles}(a) and compare the result to the model spectrum $\mathrm{rect}(\tfrac{\lambda-\lambda_{\mathrm{c}}}{\Delta\lambda})$, with $\lambda_{\mathrm{c}}\!=\!799$~nm and $\Delta\lambda\!=\!2$~nm. The marginal temporal profile $I_{t}(\tau)$, or pulse profile at the beam center, is obtained via three different approaches, all of which are in excellent agreement [Fig.~\ref{Fig:SeparateSpatialAndTemporalProfiles}(b)]: first, we obtain $I_{t}(\tau)\!=\!|\psi_{t}(\tau)|^{2}$ directly from the measured spatiotemporal intensity profile in Fig.~\ref{Fig:SpatioTemporalSpectrumAndProfile}(c), $I_{t}(\tau)\!=\!I(x\!=\!0,z\!=\!0;\tau)$; second, we calculate the Fourier transform of the measured marginal spectrum in Fig.~\ref{Fig:SeparateSpatialAndTemporalProfiles}(a); and third, we calculate the Fourier transform of the spectral model. As seen in Fig.~\ref{Fig:SeparateSpatialAndTemporalProfiles}(b), all three profiles are in excellent agreement.

\subsubsection{STWP}

Because the STWP is obtained from the reference pulse via spatiotemporal spectral-phase modulation, we expect that the marginal \textit{temporal} spectrum of the STWP to match that of the reference pulse, whereas its marginal \textit{spatial} spectrum is expected to differ fundamentally from that of the reference pulse. The plots in the second row of Fig.~\ref{Fig:SeparateSpatialAndTemporalProfiles} are for the STWP profiles corresponding to those for the reference pulse in the first row. We plot in Fig.~\ref{Fig:SeparateSpatialAndTemporalProfiles}(a) the STWP marginal temporal spectrum obtained from integrating the measured parabolic spatiotemporal spectrum in Fig.~\ref{Fig:SpatioTemporalSpectrumAndProfile}(a), and find it an excellent match to that of the reference pulse. The STWP marginal temporal profile $I_{t}(\tau)$ is obtained via the same three approaches outlined above for the reference pulse. All three are in agreement and also match the results for the reference pulse. This confirms the following conclusion: the pulse profile at the beam center of a separable reference pulse or a non-separable STWP are identical if they have the same temporal spectrum $\Delta\lambda$.

\subsection{Marginal spatial profile: Beam profile at the pulse center}

\subsubsection{Reference pulse}

The marginal spatial spectrum $\widetilde{\psi}_{x}(k_{x})\!=\!\int\!d\Omega\widetilde{\psi}(k_{x},\Omega)$ is a narrow function of width $\delta k_{x}$ for the reference pulse (a delta function in the limit of an ideal plane wave). Using the measured spatiotemporal spectrum $\widetilde{\psi}(k_{x},\lambda)$ in Fig.~\ref{Fig:SpatioTemporalSpectrumAndProfile}(a), we obtain the marginal spatial spectrum plotted in Fig.~\ref{Fig:SeparateSpatialAndTemporalProfiles}(c). For comparison, we also plot in Fig.~\ref{Fig:SeparateSpatialAndTemporalProfiles}(c) the model marginal spatial spectrum $\widetilde{\psi}_{x}(k_{x})\!\propto\!\exp\{-\tfrac{k_{x}^{2}}{2(\delta k_{x})^{2}}\}$, which provides a good match. Next, we plot in Fig.~\ref{Fig:SeparateSpatialAndTemporalProfiles}(d) the marginal spatial profile obtained via three different approaches: (1) the measured marginal spatial profile obtained from the measured spatiotemporal intensity profile in Fig.~\ref{Fig:SpatioTemporalSpectrumAndProfile}(c), $I_{x}(x)\!=\!I(x,z\!=\!0;\tau\!=\!0)$; (2) by taking the Fourier transform of the measured marginal spatial spectrum in Fig.~\ref{Fig:SeparateSpatialAndTemporalProfiles}(c); and (3) taking the Fourier transform of the spectral model. All three marginal spatial profiles are in good agreement. 

\subsubsection{STWP}

For the STWP, the marginal spatial spectrum has a distinct structure, as shown in Fig.~\ref{Fig:SeparateSpatialAndTemporalProfiles}(c), second row. The spatial bandwidth exceeds that of the reference pulse. The spatial spectrum reaches peak values symmetrically at the two extreme ends of the spectrum, and drops linearly as $k_{x}$ decreases, reaching zero at $k_{x}\!=\!0$, in contrast to the peak reached at $k_{x}\!=\!0$ for the reference pulse. Note that a portion of the spatial spectrum in the vicinity of $k_{x}\!=\!0$ is eliminated in our experimental setup via a spatial filter [SF in Fig.~\ref{Fig:Setup}].

To understand the origin of this specific spatial spectrum, we make use of the fact that each spatial frequency $k_{x}$ is associated in the ideal limit of an STWP with a single temporal frequency $\Omega$, $\widetilde{\psi}(k_{x},\Omega)\!=\!\widetilde{\psi}(\Omega)\delta(k_{x}-k_{x}(\Omega))$, with $k_{x}(\Omega)$ defined in Eq.~\ref{Eq:Parabola}. The delta function association allows us to write the identity $\tfrac{1}{2}|\widetilde{\psi}_{t}(\Omega)|^{2}\delta\Omega\!=\!|\widetilde{\psi}_{x}(k_{x})|^{2}\delta k_{x}$, which is evaluated at $\tfrac{\Omega}{\omega_{\mathrm{o}}}\!=\!\tfrac{k_{x}^{2}}{2k_{\mathrm{o}}(1-\cot\theta)}$. This identity simply reflects that the power in the (spatial) spectral interval $\delta k_{x}$ at $k_{x}$ is equal to that in the (temporal) spectral interval $\delta\Omega$ at $\Omega$, and the factor of $\tfrac{1}{2}$ reflects the fact that the parabola is double branched. That is, the power in the interval $\delta\Omega$ is split into two intervals $\delta k_{x}$ centered at $k_{x}$ and $-k_{x}$. Furthermore, the parabolic relationship between $\Omega$ and $k_{x}$ results in $\delta\Omega\!=\!\tfrac{c}{k_{\mathrm{o}}|1-\cot\theta|}|k_{x}|\delta k_{x}$. In the model for the marginal temporal spectrum we have $|\widetilde{\psi}_{t}(\Omega)|^{2}$ flat in the range from $\Omega\!=\!0$ to $\Delta\omega$, which corresponds to $k_{x}$ extending from $k_{x}\!=\!0$ to $|k_{x}|\!=\!\Delta k_{x}$, where $\tfrac{\Delta\omega}{\omega_{\mathrm{o}}}\!=\!\tfrac{(\Delta k_{x})^{2}}{2k_{\mathrm{o}}^{2}|1-\cot\theta|}$. By normalizing the temporal spectrum such that $\int_{0}^{\Delta\omega}d\Omega|\widetilde{\psi}_{t}(\Omega)|^{2}\!=\!1$, it is straightforward ro show that $|\widetilde{\psi}_{x}(k_{x})|^{2}\!=\!\tfrac{|k_{x}|}{(\Delta k_{x})^{2}}$, from which we also have a normalized spatial spectrum $\int_{-\Delta k_{x}}^{\Delta k_{x}}dk_{x}|\widetilde{\psi}_{x}(k_{x})|^{2}\!=\!1$. This analysis can be readily extended to other models for the marginal temporal spectrum. For example, if we have $|\widetilde{\psi}_{t}(\Omega)|^{2}\!\propto\!\sqrt{\Omega}$, then the marginal spatial spectrum becomes $|\widetilde{\psi}_{x}(k_{x})|^{2}\!\propto\!k_{x}^{2}$. We have found that this model provides a fit for narrow spectral bandwidths $\Delta\omega$, where the flat spectral model we utilized here is not satisfactory.


\begin{figure}[t]
    \centering
    \includegraphics[width=86mm]{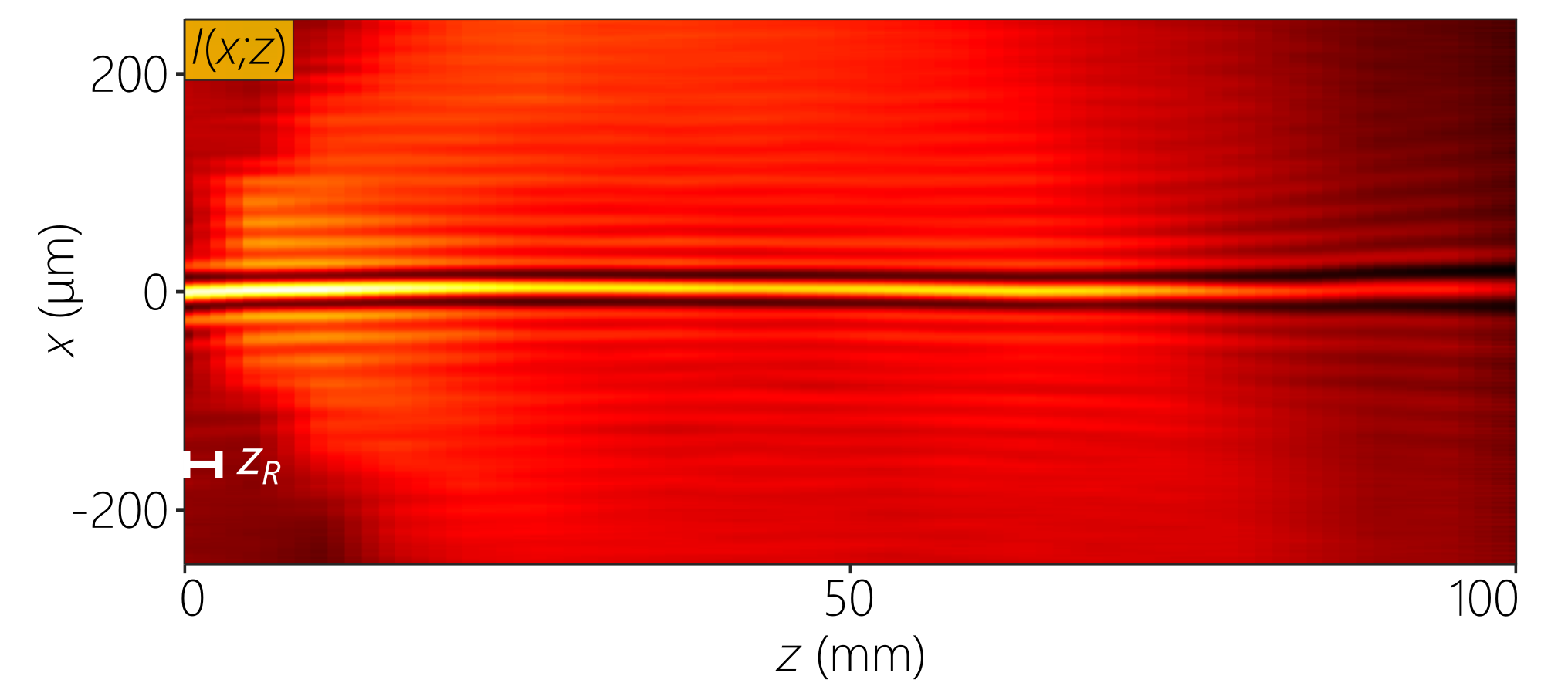}
    \caption{Measured time-averaged intensity profile $I(x,z)$ of an STWP ($\theta\!=\!50^{\circ}$), corresponding to the STWP in Fig.~\ref{Fig:SpatioTemporalSpectrumAndProfile}, obtained by scanning CCD$_{2}$ axially along $z$ after blocking the path of the reference pulse [Fig.~\ref{Fig:Setup}]. The white scale bar corresponds to the Rayleigh range $z_{\mathrm{R}}$ of a Gaussian beam whose width matches the width of the central peak of the STWP.}
    \label{Fig:Intensity}
\end{figure}

\section{Time-averaged intensity}

A hallmark of the successful synthesis of an STWP is that the time-averaged intensity $I(x,z)\!=\!\int\!d\tau\;|\psi(x,z;\tau)|^{2}$, which is the intensity recorded by a slow detector such as a CCD camera, is \textit{diffraction-free}; that is, $I(x,z)$ is constant along $z$ for a distance that far exceeds the Rayleigh range of a Gaussian beam having width matching $\Delta x$ for the STWP. We plot the measured $I(x,z)$ in Fig.~\ref{Fig:Intensity} obtained by scanning CCD$_{2}$ axially [Fig.~\ref{Fig:CharacterizationSchematics}(a) and Fig.~\ref{Fig:Setup}] while blocking the path of the reference pulse. At each axial plane $z$, the transverse intensity profile takes the form of an extended background pedestal atop of which is a spatially localized feature, as expected from Eq.~\ref{eq:TimeAveraged}.

The width of the transverse intensity feature is $\Delta x\!\approx\!26.5$~$\mu$m, which extends invariantly along the propagation axis for a distance $L_{\mathrm{max}}\!\approx\!80$~mm. In contrast, the Rayleigh range is $z_{\mathrm{R}}\!\approx\!0.7$~mm for a Gaussian beam of width 26.5~$\mu$m. The propagation distance $L_{\mathrm{max}}$ is determined by two factors: the spectral tilt angle $\theta$ and the spectral uncertainty $\delta\lambda$, which in our case is determined by the spectral resolution of the grating \cite{Yessenov19OE}.

\begin{figure}[t]
    \centering
    \includegraphics[width=86mm]{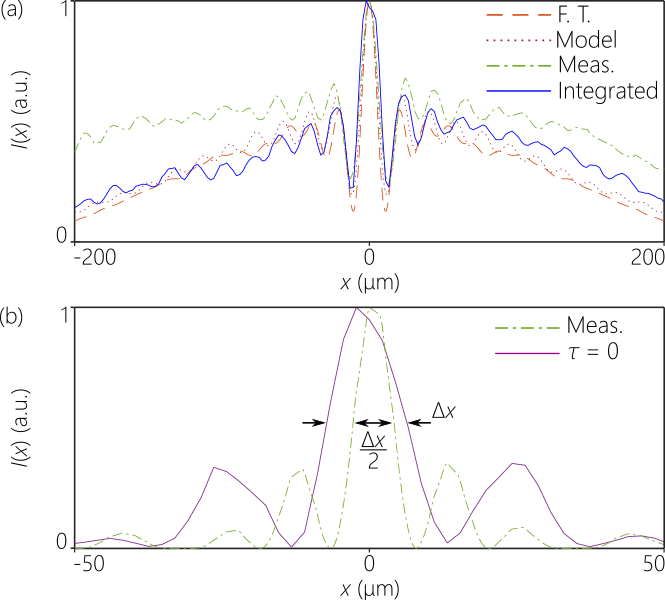}
    \caption{Time-averaged profile of an STWP at $z\!=\!0$. (a) We plot $I(x,0)$ obtained in 4 different ways: measured profile from Fig.~\ref{Fig:Intensity} (`Meas.'); by integrating the measured spatiotemporal intensity profile in Fig.~\ref{Fig:SpatioTemporalSpectrumAndProfile}(c) over $\tau$ (`Integrated'); calculated spatial profile by substituting the measured marginal spatial spectrum [Fig.~\ref{Fig:SeparateSpatialAndTemporalProfiles}(c)] into Eq.~\ref{eq:TimeAveraged} (`F.T.'); and calculated by substituting the spectral model into Eq.~\ref{eq:TimeAveraged} (`Model'). (b) The central spatial feature of the measured time-averaged intensity of the STWP from (a) after removing the background pedestal term, compared to the beam profile at the pulse center $\tau\!=\!0$ obtained from the measured spatiotemporal intensity profile in Fig.~\ref{Fig:SpatioTemporalSpectrumAndProfile}(c), second row. The transverse spatial width of the former is half of that of the latter.}
    \label{Fig:TemporalProfile}
\end{figure}

We next examine in more detail the transverse spatial structure of the time-averaged intensity. We plot in Fig.~\ref{Fig:TemporalProfile}(a) four transverse profiles at a fixed axial plane $z\!=\!0$: (1) $I(x)\!=\!I(x,0)$ obtained from the measured time-averaged intensity $I(x,z)$ plotted in Fig.~\ref{Fig:Intensity}; (2) integrating the measured spatiotemporal intensity profile in Fig.~\ref{Fig:SpatioTemporalSpectrumAndProfile}(c) over time $I(x,0)\!=\!\int\!d\tau\;I(x,0;\tau)$; (3) the spatial profile calculated via Eq.~\ref{eq:TimeAveraged} after substituting the measured spatiotemporal spectrum in Fig.~\ref{Fig:SpatioTemporalSpectrumAndProfile}(a); and (4) the calculated spatial profile using the spectral model in Fig.~\ref{Fig:SpatioTemporalSpectrumAndProfile}(b) in Eq.~\ref{eq:TimeAveraged}. We find excellent agreement with regards to the central spatial feature. Moreover, the pedestal of all these cases are in excellent agreement, except for the measured profile in which the pedestal is higher than that in the other three profiles. This is likely due to an over-estimation of the value of the spectral uncertainty $\delta\lambda$.

A crucial feature regarding the width of the time-averaged intensity profile is emphasized in Fig.~\ref{Fig:TemporalProfile}(b). We plot the time-averaged transverse intensity profile $I(x)\!=\!I(x,0)$ from Fig.~\ref{Fig:Intensity} after removing the pedestal so as to isolate the spatially localized feature at the beam center (the second term in Eq.~\ref{eq:TimeAveraged}). We compare this transverse distribution to the beam profile at the pulse center $I(x,z\!=\!0;\tau\!=\!0)$ obtained from the measured spatiotemporal intensity profile from Fig.~\ref{Fig:SpatioTemporalSpectrumAndProfile}(c), second row. As discussed earlier, for a conventional pulsed optical beam in which the spatial and temporal DoFs are separable, these two profiles are identical. In contrast, for an STWP we note in Fig.~\ref{Fig:TemporalProfile}(b) the expected discrepancy between the widths of these two profiles as a consequence of the \textit{non-separability} of the spatial and temporal DoFs. Indeed, the time-averaged intensity is narrower than the beam width at the pulse center by a factor $\approx\!2$ in accordance with the prediction in Eq.~\ref{eq:TimeAveraged}.

\begin{figure*}[t!]
    \centering
    \includegraphics[width=160mm]{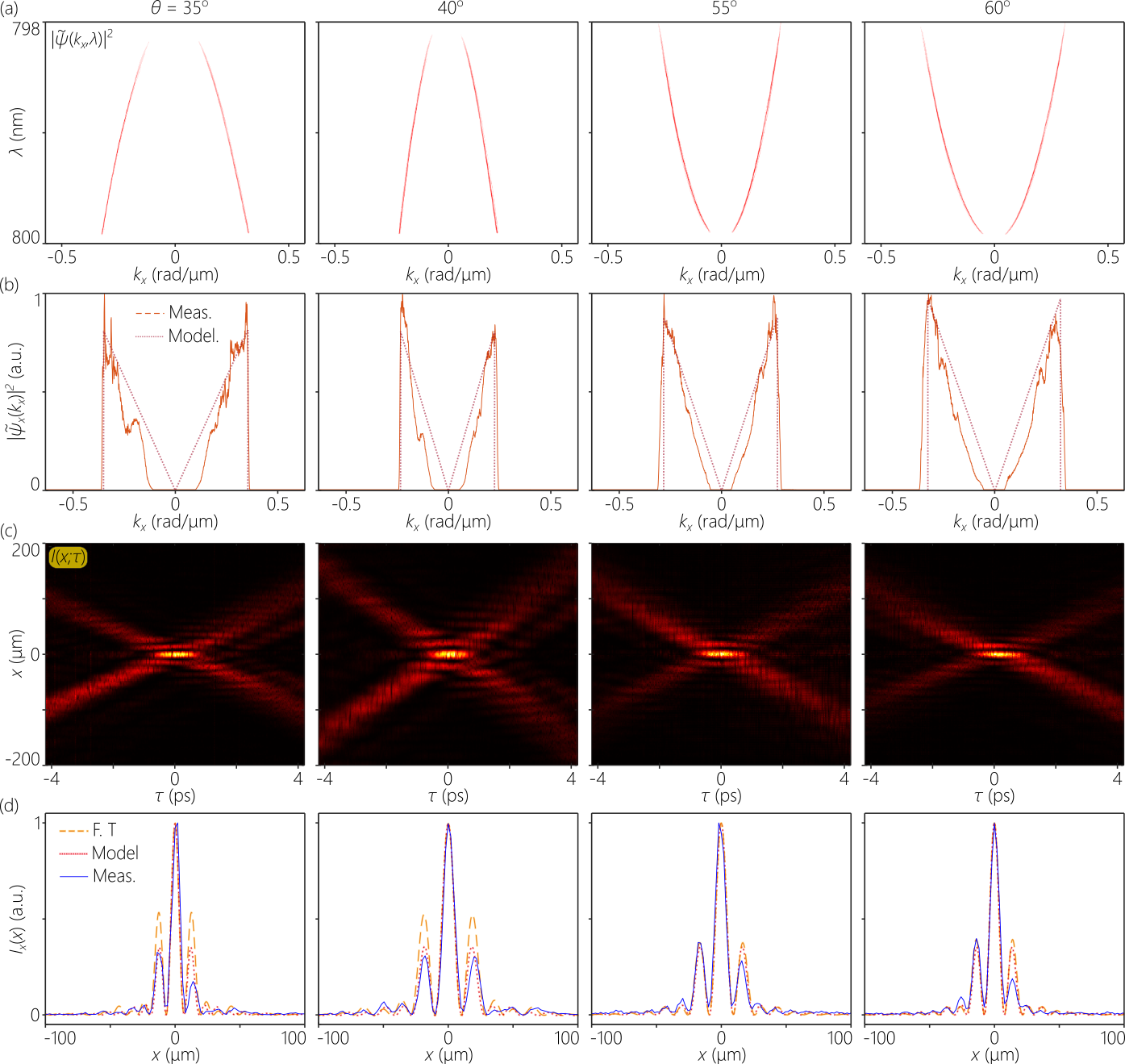}
    \caption{Tuning the spectral tilt angle $\theta$ for an STWP. The columns correspond to different values of $\theta$: subluminal STWPs with $35^{\circ}$ and $40^{\circ}$, and superluminal STWPs with $55^{\circ}$ and $60^{\circ}$. (a) Measured spatiotemporal spectra $|\widetilde{\psi}(k_{x},\lambda)|^{2}$. (b) Marginal spatial spectra $|\widetilde{\psi}_{x}(k_{x})|^{2}$ obtained by integrating the spectra in (a), compared to those obtained from the spectral model. (c) The experimentally reconstructed spatiotemporal intensity profiles $I(x,z\!=\!0;\tau)$. (d) Marginal spatial profiles $I_{x}(x)$ obtained by: taking the measured beam profile (`Meas.') at the pulse center in (c); taking the Fourier transform (`F.T.') of the measured marginal spatial spectrum in (b); and taking the Fourier transform of the spectral model (`Model'). All three are in excellent agreement for all values of $\theta$.}
    \label{Fig:DifferentTheta}
\end{figure*}

\section{Tuning the spectral tilt angle}

Thus far, all the measurements reported above have concerned an STWP at a spectral tilt angle $\theta\!=\!50^{\circ}$. As $\theta$ is tuned, the group velocity $\widetilde{v}\!=\!c\tan{\theta}$ also changes, but some other structural features of the STWP also change. In this Section, we examine the changes in the STWP structure as $\theta$ is tuned from the subluminal regime ($\theta\!<\!45^{\circ}$; here $\theta\!=\!35^{\circ}$ and $40^{\circ}$) to the superluminal regime ($\theta\!>\!45^{\circ}$; here $\theta\!=\!55^{\circ}$ and $60^{\circ}$). We hold the temporal bandwidth fixed at $\Delta\lambda\!\approx\!2$~nm. As such, the marginal temporal profile $I_{t}(\tau)$ remains invariant. However, the spatiotemporal spectra and the marginal spatial profiles undergo substantial changes.

To tune $\theta$ for an STWP, the only change to be implemented in the setup shown in Fig.~\ref{Fig:Setup} is to impart an appropriate 2D SLM phase distribution, as shown in the examples depicted in Fig.~\ref{Fig:SLMPhasePatterns}(a). We plot in Fig.~\ref{Fig:DifferentTheta}(a) the measured spatiotemporal spectra $|\widetilde{\psi}(k_{x},\lambda)|^{2}$ associated with the four STWPs at $\theta\!=\!35^{\circ},40^{\circ},55^{\circ}$, and $60^{\circ}$. Although the spectra are all parabolic in shape, we note immediately two crucial features. First, the sign of the spectral curvature changes as we traverse the luminal condition at $\theta\!=\!45^{\circ}$. That is, the sign of the curvature for the subluminal STWPs is opposite that of the superluminal STWPs. Second, the ratio of the temporal bandwidth $\Delta\lambda$ to the spatial bandwidth $\Delta k_{x}$ depends on $\theta$:
\begin{equation}
\frac{\Delta\lambda}{\lambda_{\mathrm{o}}}\approx\frac{(\Delta k_{x})^{2}}{k_{\mathrm{o}}^{2}|1-\cot{\theta|}}.
\end{equation}
Consequently, because $\Delta\lambda$ is held constant, $\Delta k_{x}$ increases as we move away from $\theta\!=\!45^{\circ}$ in either the subluminal or the superluminal directions. Whereas the marginal temporal spectrum remains the same for all $\theta$, the marginal spatial spectrum changes as shown in Fig.~\ref{Fig:DifferentTheta}(b). Although the overall structure of the spatial spectrum is invariant, the bandwidth $\Delta k_{x}$ changes, and thus also the linear slope in the spectrum varies with $\theta$. Except for the change in spectral curvature and $\Delta k_{x}$, the spectral model developed above remains valid. Estimating the spatial width $\Delta x$ of the marginal spatial profile as $\Delta x\!\sim\!\tfrac{\pi}{\Delta k_{x}}$, it is straightforward to show that: 
\begin{equation}\label{eq:AngleDependance}
\frac{\Delta x}{\lambda_{\mathrm{o}}}\approx\frac{1}{2}\frac{1}{\sqrt{\frac{\Delta\lambda}{\lambda_{\mathrm{o}}}|1-\cot{\theta}|}}.
\end{equation} 
In other words, in terms of the dependence of the width $\Delta x$ on the spectral tilt angle $\theta$, we have 
$\Delta x(\theta)\!\propto\!|1-\cot{\theta}|^{-1/2}$.

We plot in Fig.~\ref{Fig:DifferentTheta}(c) the experimentally reconstructed spatiotemporal intensity profiles $I(x,z\!=\!0;\tau)$ and in Fig.~\ref{Fig:DifferentTheta}(d) the marginal spatial profiles $I_{x}(x)$. The curves representing the theoretical expectations in Fig.~\ref{Fig:DifferentTheta} are all in excellent agreement with the measurements.

\begin{figure}[t]
    \centering
    \includegraphics[width=86mm]{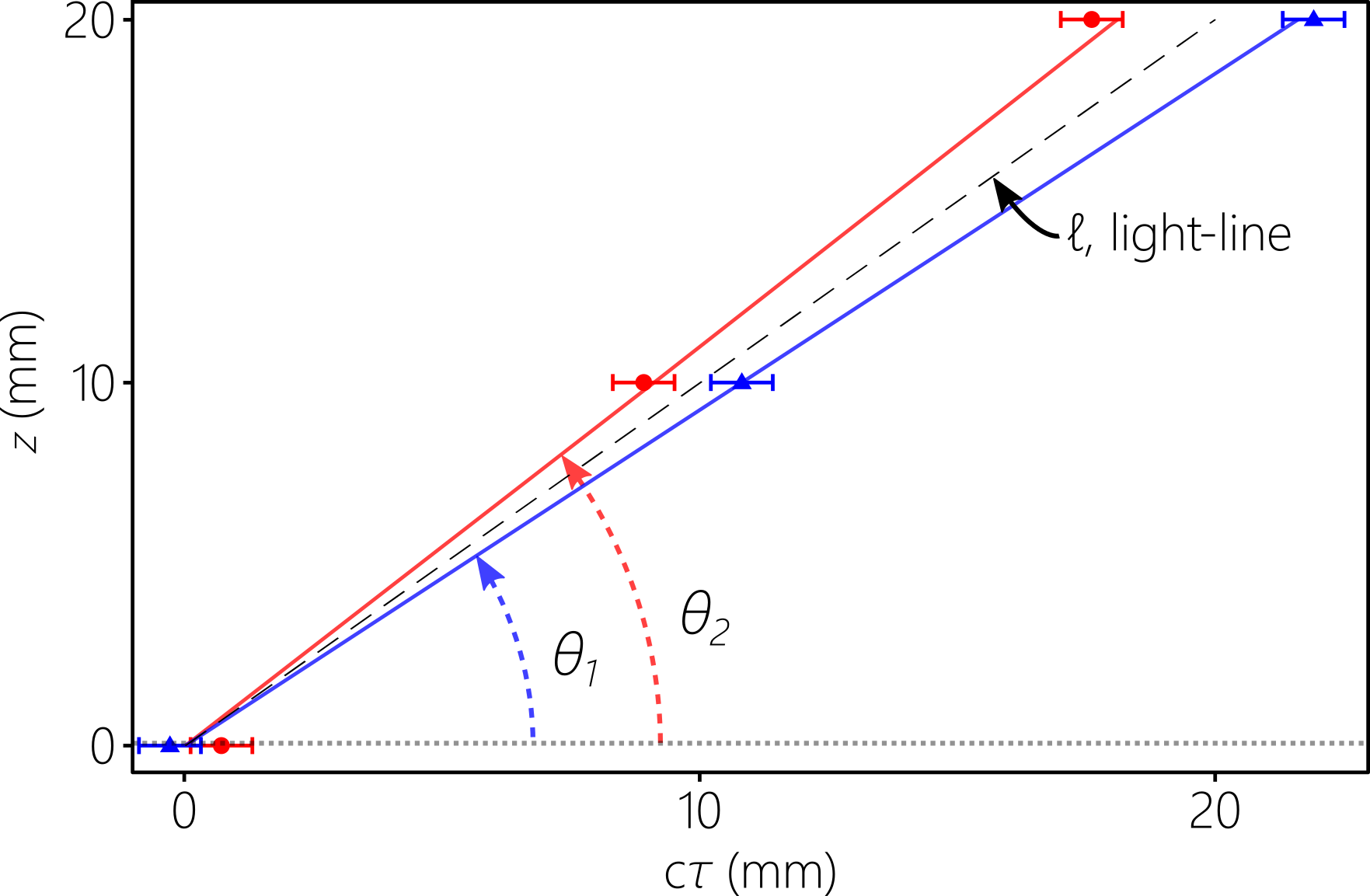}
    \caption{Group velocity measurements for STWPs with spectral tilt angles $\theta_{1}\!=\!40^{\circ}$ and $\theta_{2}\!=\!50^{\circ}$. The dashed line represents a luminal (reference) pulse with $\theta\!=\!45^{\circ}$ ($\widetilde{v}\!=\!c$) for comparison. The measured group velocities are $\widetilde{v}_{1}\!=\!0.82c$ and $\widetilde{v}_2 = 1.11c$, which correspond to spectral tilt angles of $\theta_{1}\!=\!39.25^\circ$ and $\theta_{2}\!=\!47.99^\circ$, representing errors of $1.62\%$ and $4.01\%$, respectively.}
    \label{Fig:MeasuredVelocity}
\end{figure}

\section{Measuring the group velocity}\label{sec:GroupVelocity}

One of the distinguishing features of STWPs -- in contrast to X-waves \cite{Lu92IEEEa,Lu92IEEEb,Saari97PRL}, FWMs \cite{Brittingham83JAP,Reivelt00JOSAA,Reivelt02PRE}, and other examples of propagation-invariant wave packets \cite{Reivelt03arxiv,FigueroaBook8,Turunen10PO,FigueroaBook14} -- is that their group velocity is readily tunable over a very wide range of values while remaining in the paraxial regime \cite{Kondakci19NC,Yessenov22AOP}. We described above the changes in the spatiotemporal spectra and intensity profiles of an STWP as we tune the spectral tilt angle $\theta$. A further consequence of tuning $\theta$ is that the STWP group velocity changes, $\widetilde{v}\!=\!c\tan{\theta}$. Previous measurements of the group velocity of X-waves required extremely precise techniques because the deviation of $\widetilde{v}$ from $c$ was minute, typically $\tfrac{\widetilde{v}-c}{c}\!\approx\!10^{-4}-10^{-5}$ \cite{Bowlan09OL}. Recent measurements of the group velocity of pulsed Bessel beams required precise Hong-Ou-Mandel interferometry \cite{Hong87PRL} to estimate $\widetilde{v}$ with single photons \cite{Giovannini15Science} (once again the deviation of $\widetilde{v}$ from $c$ was on the order of $10^{-5}c$). In contrast, we can estimate the group velocity of STWPs much more easily because $\widetilde{v}$ can deviate substantially from $c$ \cite{Kondakci19NC,Yessenov19OE}, by more than $4-5$ orders-of-magnitude compared to previous results.

To estimate $\widetilde{v}$, we make use of the same two-path linear interferometer shown in Fig.~\ref{Fig:CharacterizationSchematics}(b) and Fig.~\ref{Fig:Setup}. Initially, the delay $\tau$ is adjusted to overlap the synthesized STWP ($\Delta\lambda\!\approx\!2$~nm) and the reference plane-wave pulse ($\Delta\lambda\!\approx\!10$~nm) in space and time, as evident from the spatially resolved fringes observed at the CCD. By tuning $\tau$ around this value, but within the coherence time of the STWP, we reconstruct the spatiotemporal intensity profile $I(x,z\!=\!0;\tau)$. In conventional interferometric scenarios, moving the detector or CCD at the interferometer output does not impact any observations -- only changing the relative delay $\tau$ has any effect. In contrast, translating CCD$_{2}$ in Fig.~\ref{Fig:Setup} along the $z$-axis changes the measurement outcome for STWPs \cite{Yessenov19OL}. The reason is that the STWP travels at a group velocity $\widetilde{v}$ whereas the plane-wave reference pulse travels at a group velocity $c$. Therefore, displacing CCD$_{2}$ along $z$ introduces an extra group delay $\Delta\tau$ between the two wave packets. For a large enough displacement of CCD$_{2}$, the interference between the STWP and the reference pulse is completely eliminated. By changing the relative delay $\tau$ in the reference arm in the interferometer, one can compensate for that additional relative group delay and regain the interference. Using this approach, one can straightforwardly estimate $\widetilde{v}$ \cite{Kondakci19NC,Yessenov19OE}. 

We plot in Fig.~\ref{Fig:MeasuredVelocity} measurement results for two STWPs: a subluminal STWP with $\theta_{1}\!=\!40^{\circ}$ ($\widetilde{v}_{1}\!\approx\!0.82c$), and a superluminal STWP at $\theta_{2}\!=\!50^{\circ}$ ($\widetilde{v}_{2}\!\approx\!1.11c$). In both cases, we plot the additional relative delay $\tau$ needed to overlap the STWP and the reference pulse as CCD$_{2}$ is displaced along $z$. The results lie along a straight line whose slope provides a direct estimate of $\widetilde{v}\!=\!c\tan{\theta}$. Indeed, the straight-line fit in the plot in Fig.~\ref{Fig:MeasuredVelocity} makes an angle with the horizontal that corresponds to the spectral tilt angle $\theta$.

\section{Discussion and Conclusions}

We have described in detail the setup for synthesizing STWPs in the form of light sheets, along with the necessary characterization procedures for benchmarking successful synthesis of these unique optical-field configurations. We hope that this detailed description suffices for any newcomer to set up their own system for synthesizing and characterizing STWPs. The linear synthesis system utilizes a phase-only spatiotemporal Fourier synthesis strategy, and this precisely controllable approach is thus energy-efficient. By endowing the field with programmable spatiotemporal spectral structure, we can tune the group velocity of the STWP continuously and smoothly -- with no moving parts -- away from $c$ in free space over a broad range of values (from subluminal to superluminal and even negative-$\widetilde{v}$ regimes), while remaining in the paraxial regime, simply by implementing the appropriate phase distribution imparted to the spectrally resolved wave front. 

We have focused on preparing propagation-invariant STWPs traveling at a fixed group velocity $\widetilde{v}$. Nevertheless, the same setup can be employed in producing a wide variety of other classes of STWPs by simply changing the phase distribution imparted by the SLM to the spectrally resolved wave front (i.e., modifying the SLM phase patterns in Fig.~\ref{Fig:SLMPhasePatterns}). These include axially accelerating STWPs \cite{Clerici08OE,ValtnaLukner09OE,Yessenov20PRL2,Li20SR,Li20CP,Li21CP,Hall22OLArbAccel}, axially encoded spectra \cite{Motz21PRA}, STWPs that incur group-velocity dispersion in free space \cite{Yessenov21ACSP,Hall21PRA,Hall21OLNormalGVD} and are thus dispersion-free in dispersive media \cite{Porras01OL,Malaguti08OL,Malaguti09PRA,Yessenov22OLdispersion,He22LPR,Bejot22ACSP,Hall23LPR,Hall23Normal,Hall23NPhys}, STWPs that are circularly symmetric in time and space while traveling in the anomalous dispersion regime \cite{Hall23NPhys} (the so-called Mackinnon wave packet \cite{Mackinnon78FP}), and even tests of the relativistic transformations of STWPs \cite{Belanger84JOSAA,Saari04PRE,Ramsey23PRA,Yessenov23PRA,Yessenov23arxiv}. Moreover, using the techniques described here, it is possible to explore new potential applications in novel optical waveguide and fiber modal structures \cite{Vengsarkar92JOSAA,Zamboni01PRE,Zamboni02PRE,Zamboni03PRE,Shiri20NC,Kibler21PRL,Guo21PRR,Ruano21JO,Bejot21ACSP,Shiri22ACSP,Bejot22ACSP,Stefanska23ACSP,Jolly23JO,Shiri23JOSAA}, broadband interactions with optical cavities \cite{Shiri20OL,Shiri20APLP}, and surface-plasmon polaritons \cite{Schepler20ACSP,Diouf22AO,Ichiji23ACSP,Ichiji24JOSAA}, among other emerging possibilities \cite{Kondakci18OL,Diouf21OE,Diouf22SR,Diouf23Optica}.

In conclusion, we have presented a detailed theoretical model for STWPs in the form of light sheets. In addition, we have described an experimental apparatus for synthesizing such STWPs and for characterizing their unique spatiotemporal structure in both the spectral and physical domains. Spatiotemporal structured optical fields represent a new frontier in optical physics, whereby exotic propagation characteristics and novel interactions with photonic devices are made possible. We hope that this tutorial can help launch other researchers along new directions in this rapidly developing field.

\textbf{Funding}
U.S. Office of Naval Research (ONR) N00014-17-1-2458 and N00014-20-1-2789.

\noindent
\textbf{Disclosures}
The authors declare no conflicts of interest.

\noindent
\textbf{Data availability}
Data underlying the results presented in this paper are not publicly available at this time but may be obtained from the authors upon reasonable request.

\bibliography{diffraction}

\end{document}